\documentclass[useAMS,usenatbib, fleqn]{mnras}
\usepackage{ae,aecompl}
\usepackage[dvipsnames]{xcolor}

\title[Clusters in different density spiral arm environments]
{The formation of clusters and OB associations in different density spiral arm environments
}
\author[Dobbs]
{C. L. Dobbs\thanks{E-mail:
C.L.Dobbs@exeter.c.uk}$^{1}$, T. J. R. Bending$^{1}$, A. R. Pettitt$^{2}$, A. S. M. Buckner$^{1}$, M. R. Bate$^{1}$\\
$^{1}$School of Physics and Astronomy, University of Exeter, Stocker Road, Exeter, EX4 4QL, UK\\
$^{2}$Department of Physics and Astronomy, California State University, Sacramento, 6000 J Street, Sacramento, CA 95819-6041, USA\\
}
\usepackage{amssymb}
\usepackage{amsmath}
\usepackage[pdftex]{graphicx}
\usepackage{epsfig}
\usepackage[flushleft]{threeparttable}
\def\aap{{ A\&A}}

\def\apjl{{ApJL}}

\def\apjs{{ApJS}}

\begin{document}
\label{firstpage}
\date{\today}

\pagerange{\pageref{firstpage}--\pageref{lastpage}} \pubyear{2022}

\maketitle

\begin{abstract}
We present simulations of the formation and evolution of clusters in spiral arms. The simulations follow two different spiral arm regions, and the total gas mass is varied to produce a range of different mass clusters. We find that including photoionizing feedback produces the observed cluster mass radius relation, increasing the radii of clusters compared to without feedback. Supernovae have little impact on cluster properties. We find that in our high density, high gas mass simulations, star formation is less affected by feedback, as star formation occurs rapidly before feedback has much impact. In our lowest gas density simulation, the resulting clusters are completely different (e.g. the number of clusters and their masses) to the case with no feedback. The star formation rate is also significantly suppressed. The fraction of stars in clusters in this model decreases with time flattening at about 20\%. In our lowest gas simulation model, we see the formation of a star forming group with properties similar to an OB association, in particular similar to Orion Ia. We suggest that low densities, and stronger initial dynamics are conducive to forming associations rather than clusters. In all models cluster formation is complex with clusters merging and splitting. The most massive clusters which form have tended to undergo more mergers.  
\end{abstract} 


\section{Introduction}

Whilst numerous studies now follow cluster formation on molecular cloud, and even galaxy scales, we still do not yet have a good understanding of how different types of clusters or stellar groups form, under what conditions and what sets the properties of stellar clusters.
Groups of stars can be categorised broadly as globular clusters, massive clusters, open clusters, OB associations, T associations and moving groups. In our own Milky Way Galaxy, young massive clusters, characterised by masses $\gtrsim10^4$ M$_{\odot}$ \citep{PZ2010} and usually small age spreads ($\sim$ Myr) \citep{Longmore2014}, appear to be rare, with only a few examples away from the Galactic Centre \citep{Joshi2016}. Most ongoing sites of star formation appear to be lower mass, likely open clusters or larger more dispersed associations. For example, in \citet{PZ2010}, 8 clusters are shown as lying in the parameter space of YMCs (Fig. 2) compared to 10’s of open clusters with masses up to a few 1000 M$_{\odot}$. For external galaxies, studies of clusters, e.g. mass, age functions, are often limited to $\gtrsim5000$ M$_{\odot}$ (e.g. \citealt{Bastian2012,Chandar2014,Mulia2016, Messa2018,Krumreview2019}), except for Local Group galaxies (e.g. \citealt{Johnson2017}). However using the recent LEGUS \citep{Calzetti2015} survey, \citet{Brown2021} study over 6000 clusters from many nearby galaxies down to lower masses, including unbound associations, although they find LEGUS still preferentially selects bound clusters.

\citet{Elmegreen1977} supposed that larger associations of stars were produced by consecutive feedback events which triggered successive star formation events, which would explain the relatively large age spreads and multiple episodes of star formation observed (e.g. \citealt{Zari2019}). \citet{Elmegreen1985} proposed that such associations, rather than bound clusters, could arise in low mass clouds, or smaller quiescent regions of higher mass clouds. Following \cite{Lada2003}, associations were supposed to be remnants of bound clusters, following gas expulsion, stellar evolution and / or N-body interactions \citep{Goodwin2006,Bastian2008,Pfalzner2013}. However recent kinematic studies of associations suggest there is no simple picture of expansion \citep{Wright2016,Melnik2020,Ward2020,MApellaniz2020}. Rather the structure of associations is set by the structure of the gas when they are formed \citep{Wright2018,Goul2018,Lim2019,Ward2018}. Simulations to trace back currently observed associations suggest that they originate from sub-virial and highly structured initial conditions \citep{Farias2018,Schoettler2019}.

Numerical simulations of galaxies have generally tended to follow higher mass clusters rather than smaller clusters or associations. Again this is a consequence of resolution, and the time the clusters are evolved for. On galaxy scales, it is difficult to fully resolve clusters. Simulations of dwarf galaxies resolve the formation and evolution of clusters following the mass of stars or clusters down to the mass of massive stars \citep{Hu2016,Emerick2018,Lahen2019,Lahen2020,Hislop2022}. \citet{Hislop2022} find that at higher star formation rates, some clusters disperse, although they are not able to form realistic clusters $<1000$ M$_{\odot}$.

On smaller scales, \citet{Grudic2021} perform simulations of isolated, massive, bound, turbulent GMCs and find that they typically evolve to form a large number of small bound clusters which reside in a larger association, along with many unbound stars. \citet{Hajime2021} find that unless the star formation efficiency is high, individual clusters tend to disperse. Ionising feedback leads to less dense clusters at lower surface densities, but has little effect at high surface densities where globular clusters or YMCs are likely to form \citep{Hajime2022}.

Isolated cloud simulations however may miss external gas dynamics, such as continuing gas inflow which may continue to produce star formation. \citet{Bending2020} found that ionising radiation propagating into the surrounding medium can also trigger further star formation. Furthermore even on molecular cloud scales, it is difficult to resolve stars fully down to brown dwarfs, limiting the total mass of such simulations to around $10^4$ M$_{\odot}$. Simulations on molecular cloud scales also don't necessarily have the capacity to follow the clusters for long enough to see whether they would likely stay as open clusters, or disperse (e.g. \citealt{Bate2003,Bonnell2003}).

Simulations can take into account of surrounding gas by resimulating, or zooming in on small sections of whole galaxy simulations \citep{vanloo2013,Smilgys2017,Bending2020,Ali2021,Rieder2022,Smith2020,Dobbs2022}. \citet{Dobbs2022} showed that colliding flows on larger scales are significant in determining the final cluster mass, partly because they lead to more mergers between clusters. Generally, simulations  show that at least for the formation of more massive clusters, cluster formation is hierarchical and mergers are frequent \citep{Bonnell2003,Smilgys2017,Fujii2015,Fujii2021a,Guszejnov2022,Dobbs2022}.

To fully model clusters, one possibility for cosmological or galactic simulations is to include the cluster evolution separately using a theoretical or empirical model \citep{Pfeffer2018} or by separately evolving a sample of clusters \citep{Rodriguez2022}. An alternative approach is to model the full population of stars with Nbody dynamics \citep{Fujii2021a,Rieder2022,Liow2022}, even if the gas resolution is low, thus achieving the same resolution in the stars as individual cluster simulations. This has the advantage that the cluster dynamics can be followed explicitly. However the resolution of the stellar component would still be limiting on larger galaxy scales as modelled for longer periods of time.

In this paper, we simulate cluster formation and evolution in spiral arm regions, taken from global spiral galaxy simulations, with photoionizing and supernovae feedback. We simply include sink particles when star formation occurs, so are unable to resolve the full stellar population and dynamics, but we do follow the clusters rather than including any subgrid description of cluster evolution.
We look at the effect of feedback on the star formation rate for the different regions, which are characterised by different masses (Section~\ref{sec:sfr}). We compare the evolution of clusters as determined by finding clusters at each time frame (Section~\ref{sec:cluster_time}), and by following constituent cluster sink particles (Section~\ref{sec:partevol}). We compare cluster properties with and without feedback (Section~\ref{sec:properties}) and also compare the outcomes of our models to known OB associations (Section~\ref{sec:OBassociations}). In Section~\ref{sec:supernovae} we examine the impact of including supernovae, versus ionization only.

\section{Method}
In \cite{Dobbs2022}, we performed simulations of two sections of a Milky Way-like spiral galaxy from \citet{Pettitt2015}. In that paper we denoted these sections Regions 1 and 2, where Region 1 exhibited strongly converging flows. We concluded that the velocities, and gas densities, in particular for Region 1 of those simulations represented the most extreme environment from the initial galaxy scale simulation and so resulted in very massive clusters. Even in our second region, Region 2, the gas densities were still high compared to typical Milky Way densities. In this paper we extend the work of \citet{Dobbs2022}, by rerunning the regions presented in the previous work but with lower initial gas masses. Otherwise we keep the physics included in the simulations exactly the same as in previous work, so the only parameter which is varied is the gas mass. We could have rerun the original galaxy simulations with a lower gas mass, however, as well as avoiding running multiple whole galaxy simulations, the approach here also allows us to compare simulations where the initial conditions still reflect large scale processes such as spiral arms but are the same except for the initial densities. Rerunning the whole galaxy would likely substantiate the conclusions from our different density runs, as for example the gas would likely also be less concentrated in the midplane, with less cold gas, further increasing the difference between lower and higher density models.
\begin{table*}
\begin{tabular}{c|c|c|c|c|c}
 \hline 
Model & Region  & Gas mass & Feedback & Time evolved & Time of first\\
 & & ($10^5$ M$_{\odot}$) & & (Myr) & supernova (Myr) \\
\hline
M1R1 & 1 & 1 & N & 40 & -  \\
M1R1FB & 1 & 1 & Y & 40 & 6.26 \\
M5R1 & 1 & 5 & N & 4 & - \\
M5R1FB & 1 & 5 & Y & 4 & (4.73) \\
M5R2 & 2 & 5 & N & 6.5 & - \\
M5R2FB & 2 & 5 & Y & 6.5 &  5.1 \\
M25R1 & 1 & 25 & N & 1.2 & - \\
M25R1FB & 1 & 25 & Y & 1.2 & (4.13) \\
M25R2 & 2 & 25 & N & 4.5 & - \\
M25R2FB & 2 & 25 & Y & 4.5 & 4.35\\
\hline
\end{tabular}
\caption{Table showing the different simulations presented. The lower four were all shown in \citet{Dobbs2022}. As indicated in the Table, some simulations run past where supernovae occur, whilst others do not reach the point where supernovae start to occur.}
\label{tab:simulations}
\end{table*}

\subsection{Details of simulations}
The physics included in the simulations is essentially the same as \citet{Dobbs2022}, except for the inclusion of supernovae, but we include a summary of the previous methods here as well. We use the Smoothed Particle Hydrodynamics (SPH) code sphNG to carry out our simulations. The code is based on an original version by \citet{Benz1990}, but has since been substantially modified to include sink particles, cooling and heating, and stellar feedback. We model the whole stellar disc of the galaxy using star particles, but only include gas particles in the regions we select. The halo of the galaxy is represented by an NFW potential \citep{NFW1997}. For the gas we include self gravity, and heating, cooling, H$_2$ and CO chemistry according to \citet{Dobbs2008}, originally from \citet{Glover2007}. Sink particles are inserted once gas reaches a certain density, and the gas is converging and bound according to \citet{Bate1995}. We use the same sink criteria and parameters as the Set 2 (highest resolution) simulations from \citet{Dobbs2022}. Sinks are created for densities $>1000$ cm$^{-3}$ if they meet the above criteria, but similarly to \citet{Dobbs2022}, we enforce sink creation if densities reach $10^5$ cm$^{-3}$. The sink radius is set to 0.1 pc, the merger radius is 0.001 pc. Sink masses can be of order 10 M$_{\odot}$ in our lowest mass simulation, thus potentially representing a single very massive star, but more likely a number of low mass stars. In our highest mass simulations (i.e. M25R1, M25R1FB, M25R2 and M25R2FB shown in \citealt{Dobbs2022}), the sinks typically represent 100's of solar masses and therefore sample a full IMF of stars.

\subsubsection{Stellar feedback}
Once sinks are formed, they are allocated a population of stars according to a Kroupa \citep{Kroupa2001} Initial Mass Function (IMF). Stars above a given mass are treated as ionising sources. The photoionisation scheme for our models is the same as described in \citet{Dobbs2022} and \citet{Bending2020}. The ionisation is calculated using a line of sight method, which determines the change in ionisation fraction for every gas particle which lies within a given distance (see below) of each sink particle. Column densities are computed by summing over all particles whose smoothing length overlaps with the line of sight.
Specifically, the ionisation fraction is evolved as
\begin{equation}
\begin{aligned}
\frac{{\rm d}H_{\rm II}}{{\rm d}t} = \frac{h^2}{r^2} \left( \frac{Q_{\rm H}}{4 \pi} \right. &  -  \int^{r}_{0} r^{\prime 2} n(r^\prime)^2 \alpha_{\rm B} {\rm d}r^\prime  \\ 
& -  \left. \frac{1}{\delta t} \int^{r}_{0} r^{\prime 2} n(r^\prime) [1-H_{\rm II}(r^\prime)] {\rm d}r^\prime  \right),
\label{eq:evol_ionisation}
\end{aligned}
\end{equation}
 where $h$ is the smoothing length, $Q_{\rm H}$ is the ionising flux, $n$ is the number density, $\alpha_B$ is the recombination efficiency (here $2.7\times 10^{-13}$ cm$^3$ s$^{-1}$), $\delta t$ is the time interval and \ion{H}{II} is the ionisation fraction of the gas (from 0 to 1). Gas which is ionised is heated to a temperature of $10^4$ K. As in \citet{Dobbs2022}, the ionisation is only calculated to a certain distance, to save computational time, and the same as for Dobbs et al. 2022, this distance is roughly half the size of the simulated region. As in \citet{Dobbs2022}, we use an efficiency of 50\% such that half the mass of sinks is converted to stars, which correspondingly sets the number of massive stars undergoing feedback. Unlike \citet{Dobbs2022}, we include this efficiency when working out stellar and cluster masses; this efficiency reflects that not all gas will be turned into stars on the scales we model, and means that when we compare the masses of clusters with how many massive stars they contain this will be consistent. However there is an inconsistency in that 50\% of gas should ideally be returned to the ISM but it is retained in sink particles. 

In addition to ionisation, we also include supernovae. Supernovae are inserted the same way as described in \citet{Bending2022}, which is similar in turn to \citet{Dobbs2011}. We insert supernovae for stars >18 M$_{\odot}$. For each star >18 M$_{\odot}$ we determine the star's lifetime using the SEBA program \citep{PZ2012, Toonen2012}.  Once stars reach the end of their lifetimes, they are assumed to undergo supernovae. The energy of the supernovae is inserted as both kinetic and thermal energy and follows a snowplough solution. Supernovae do not occur until late on in our simulations, and some simulations do not reach the point where supernovae occur. The time of the first supernova, and the total evolution time for the simulations are shown in Table~\ref{tab:simulations}. Winds are not included, however we found in \citet{Ali2022} that winds have little effect in our simulations.

We note that in our sampling and feedback prescription, sinks may exhibit masses which are much less than that required to resolve the IMF. However as described in \citet{Bending2020}, we presample the stellar masses, similar to \citet{Geen2018}, and use the total mass of the simulation to assign stars of a given mass to sink particles (according also to the size of the sink particles). As such the IMF is sampled from the collective mass, which does sample the IMF. \citet{Liow2022} propose a method whereby sinks are grouped together to assign stars, which we don't use here; our method is more similar to the `All grouped' method in their paper, which well reproduces the IMF.

\subsubsection{Initial conditions}
We take our initial conditions from \citet{Pettitt2015}  as described in \citet{Dobbs2022}. In the latter paper we performed two successive resimulations of regions in the galaxy such that we obtained a resolution suitable to resolve clusters.  We selected two regions, one associated with two merging or joining spiral arms where the densities and converging velocities are high (Region 1) and another from simply a typical region along a spiral arm (Region 2) with moderate converging velocities. For both these regions, the gas densities were fairly high and massive clusters were formed. Here we run simulations with lower gas densities for comparison. We run a model of each Region, with 5 x lower gas mass, and we also run a further model of Region 1 with 25 times lower gas mass. We list these different simulations in Table~\ref{tab:simulations}. The models M25R1FB, M25R1, M25R2FB and M25R2 are those presented in \citet{Dobbs2022}, whilst the other 6 simulations (which we primarily focus on) are presented here for the first time.

The initial conditions for the new Region 2 simulations (M5R2 and M5R2FB) are exactly the same as those used in \citet{Dobbs2022}, so we can compare these directly with M25R2FB and M25R2. For the Region 1 simulations, the initial conditions from \citet{Dobbs2022} already contained sink particles, so we re-ran the first resimulation (denoted R1z85sI in \citealt{Dobbs2022}) with 5 times lower mass, and used these as initial conditions for the Region 1 simulations presented here (M1R1, M1R1FB, M5R1, M5R1FB). Unlike \citet{Dobbs2022}, the initial conditions for the Region 1 simulations presented here contain no initial sink particles. All the new Region 1 simulations here start with the same initial conditions, except that the particle masses differ. The initial conditions will be slightly different from M25R1 and M25R1FB, but we only use these previous simulations for Region 1 to compare the star formation rates, the rest of the analysis does not involve those simulations. In all simulations we include 1.1 million particles to model the stellar disc. The initial number of gas particles for the Region 1 and Region 2 simulations are  6126990 and 5025840 respectively. The mass of the particles is 0.1 M$_{\odot}$ in the models where we decrease the mass by a factor of 5 (M5R1, M5R1FB, M5R2 and M5R2FB), and 0.02 M$_{\odot}$ in M1R1 and M1R1FB.

\begin{figure}
\centerline{\includegraphics[scale=0.42]{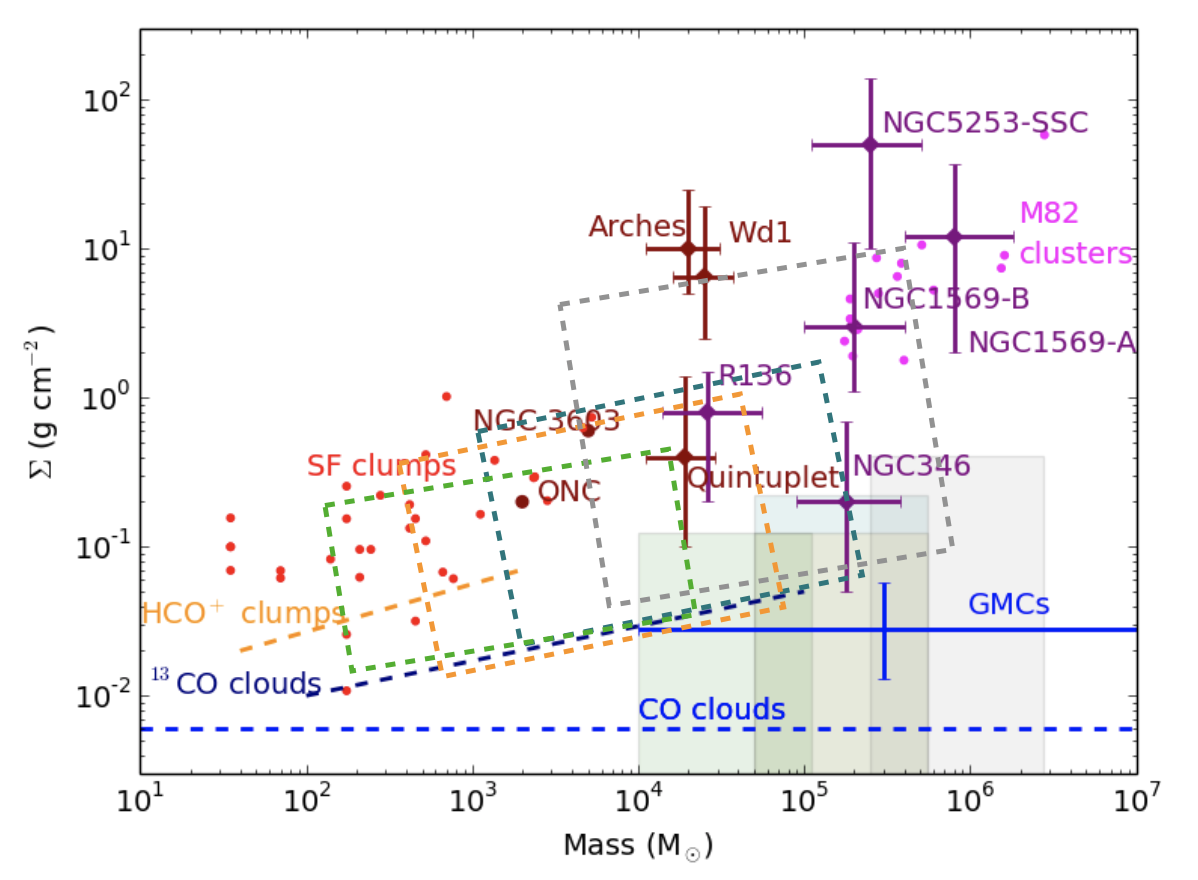}}
\caption{This figure is based on figures from previous works \citep{Tan2002,Tan2013} which show surface density plotted against mass for clusters, clumps and clouds. Here we show some of the same data shown in the previous figures, including smaller star forming clumps from \citet{Mueller2002} (red), clusters from M82 (pink) \citep{McCrady2007}, and some massive clusters from the Milky Way (dark red) and the LMC (purple) which are individually labelled. Gas densities from the simulations discussed in this paper are shown as filled boxes, and the approximate range of cluster masses and densities are shown as dashed boxes. The colours indicate the M1R1FB (green), M5R2FB (orange), M5R1FB (teal) and M25R2FB (grey) simulations. }  
\label{fig:tanfigure}
\end{figure}

In Figure~\ref{fig:tanfigure} we indicate where our simulations would lie in a plot of surface density versus mass, which is a simplified version of the figure shown in \citet{Tan2013}. We only used the simulations with feedback for this figure. The filled panels show the range of gas surface densities in the simulations, calculated at times midway through when stars are forming and there is still dense gas in the vicinity of the clusters. The width in the $x$ direction is simply one order of magnitude width up to the total gas mass in the simulation. The dashed box regions show the typical range of cluster masses and surface densities. As indicated by Figure~\ref{fig:tanfigure}, our high mass simulation M25R2FB encompasses massive clusters in the LMC, Milky Way and M82, such as R136 and Quintuplet. The lowest mass simulation (M1R1FB) lies in the same regime as the ONC, and lower mass clusters inidicated by the SF clumps.

\section{Results}

\subsection{Evolution of models}
\begin{figure*}
\centerline{\includegraphics[scale=0.86]{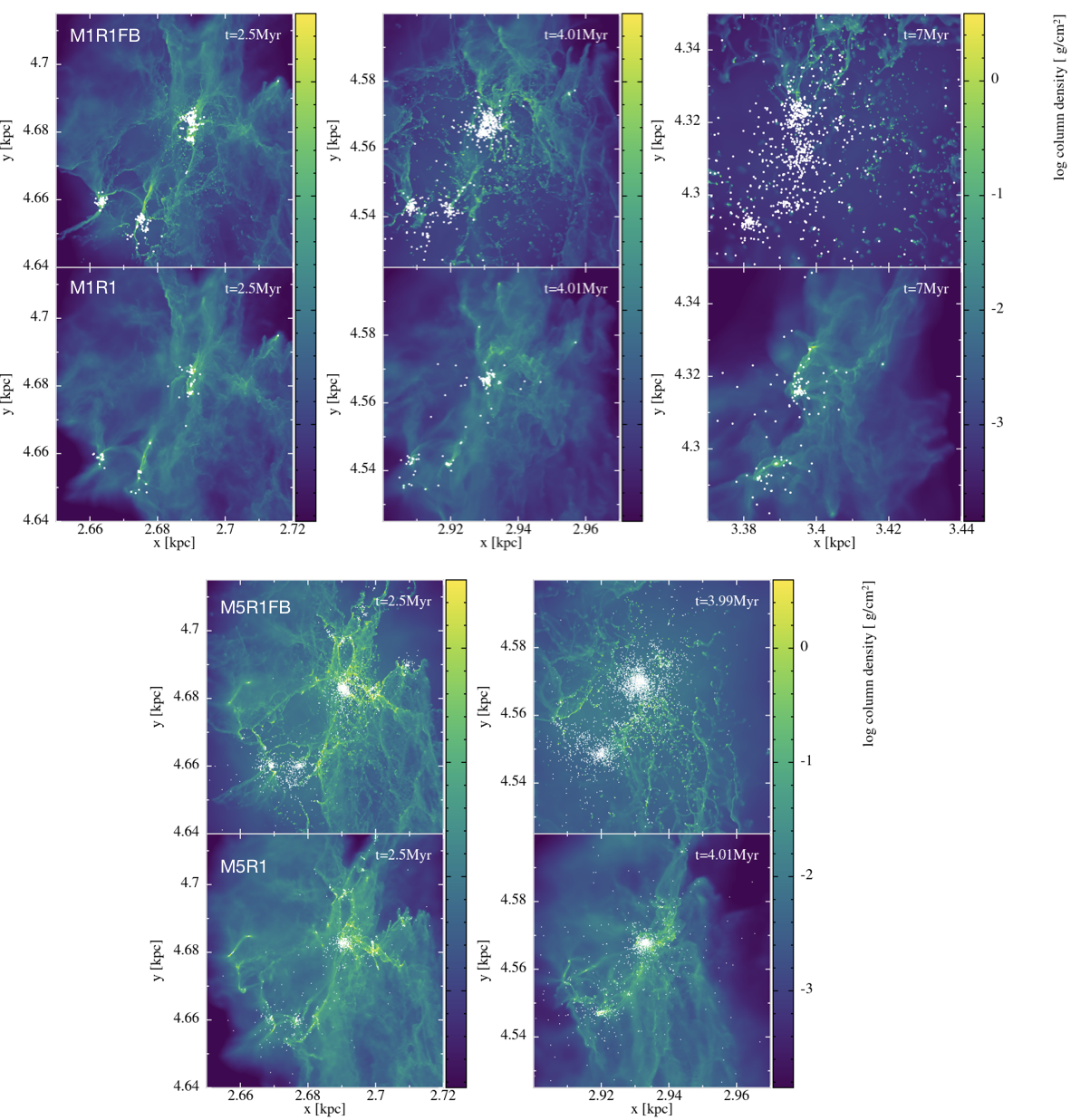}}
\caption{The evolution of the low mass (top panels) and medium mass (lower panels) Region 1 simulations are shown with and without feedback. With feedback, the photoionization is clearly producing emptier regions, and diffuse gas, and also larger more widely dispersed clusters.}  
\label{fig:region1}
\end{figure*}

\begin{figure}
\centerline{\includegraphics[scale=0.6]{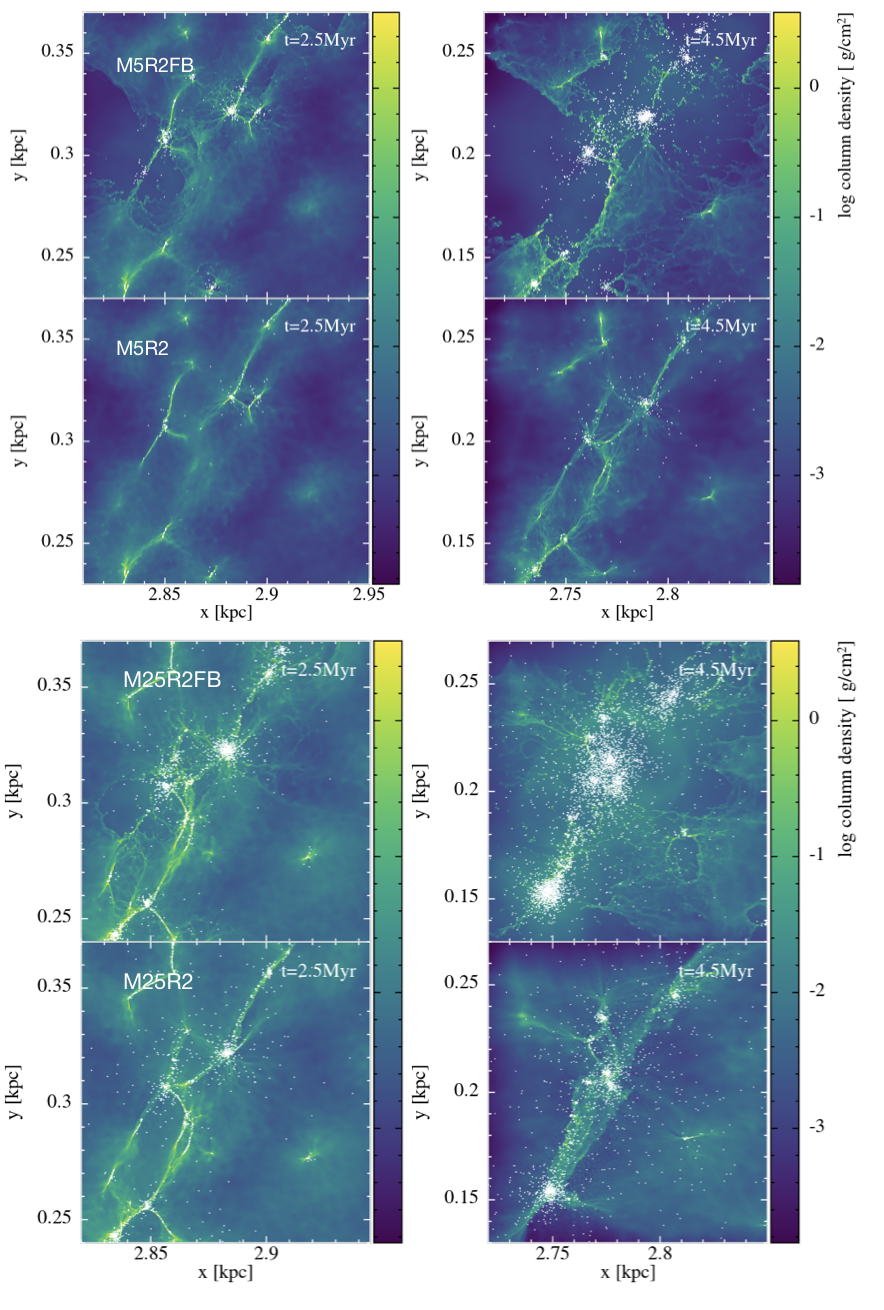}}
\caption{The evolution of the medium mass (top panels) and high mass (lower panels) Region 2 simulations are shown with and without feedback. Again photoionization is dispersing gas or leading to diffuse gas around the clusters, and leading to more spread out clusters with larger radii.}  
\label{fig:region2}
\end{figure}
We show the evolution of the Region 1 simulations in Figure~\ref{fig:region1}, the lowest gas mass (M1R1 and M1R1FB) simulations are presented in the top panels and the moderate gas mass simulations (M5R1 and M5R1FB) in the lower panels. In each case the model with feedback included is shown on the top.
For both we see that feedback has a significant effect on both the gas and the distribution of stars. For example if we look at the middle (t$=4$ Myr) panels for the low gas mass simulations, with feedback included we see large regions which are relatively devoid of gas, whilst the remaining gas is arranged into fairly dense, sharp features. By contrast without feedback, the gas is more uniformly distributed. At first glance, the ionization appears to have a greater effect in the lowest mass simulation. However if we compare at an equivalent time, e.g. the 4 Myr panels, with the low mass (top panels) the ionization produces regions devoid of very low gas density, whereas at the medium mass (lower panels), the ionization is producing large amounts of low density diffuse gas, but in both cases the ionization is having a strong effect on the gas morphology. In the low gas mass model, by 7 Myr, ionization has largely emptied the regions around the clusters of gas. Gas is present further away, so the rest of the evolution of the cluster is largely due to the N-body dynamics of the sinks.

The main difference we see in the stars is that the clusters appear larger (i.e. over a larger size scale) and more dispersed in the models with feedback compared to those without. Particularly with the medium gas mass models, the clusters cover a larger area. We also see for the low gas models (top panels) that at 7 Myr there are around 4 apparent groups of stars which are quite dispersed, whereas in the no feedback case we see three or possibly four very concentrated small clusters.

In Figure~\ref{fig:region2} we show the evolution of the Region 2 simulations, showing the medium gas mass runs (M5R2 and M5R2FB) and the high gas mass models (M25R2 and M25R2FB), again both are before supernovae occur. Again we see that the clusters are spatially much more extended with feedback. We also see the impact of the ionization producing very low density regions in the medium mass models (top) and creating large amounts of diffuse gas in the high mass models. We also see a difference between the medium and high gas mass runs. At 2.5 Myr for the medium gas mass case, there are separate clear filaments in which clusters are forming. At 4.5 Myr, these separate filaments and clusters are still apparent. However in the high mass models (M25R2 and M25R2FB), these filaments have more or less merged into one single structure. So in the higher mass model, the gravity from the higher gas mass, as well as the initial gas velocities, is producing a different morphology and clusters which are spatially closer together, and which are likely in the process of merging together.

\subsection{Star formation rates}
\label{sec:sfr}
\begin{figure}
\centerline{\includegraphics[scale=0.42]{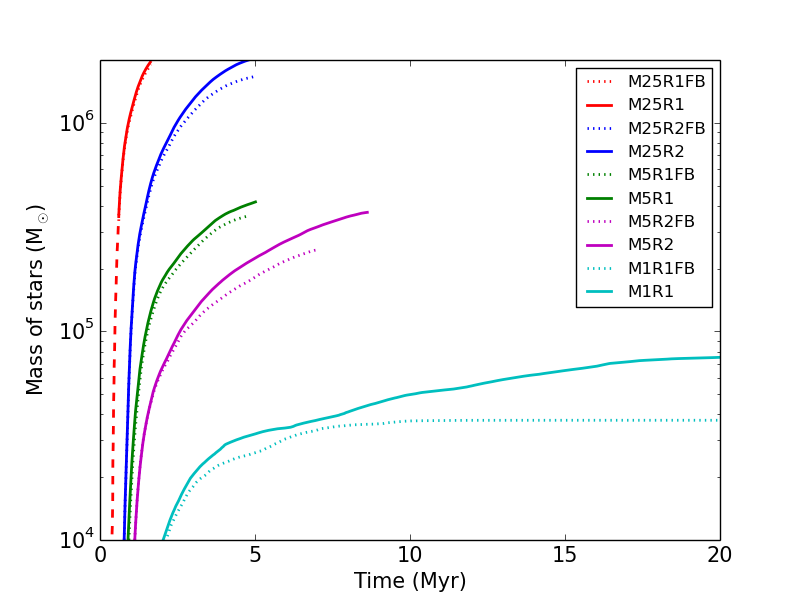}}
\caption{The mass of stars formed is shown versus time for all the different simulations. In all cases photoionization reduces the amount of stars formed, more so in the lowest mass model, M1R1FB. The red dashed part indicates evolution from a larger scale simulation for that particular cluster, as described in \citet{Dobbs2022}.}  
\label{fig:massofstars}
\end{figure}

We show the mass of stars formed versus time for the different models in Figure~\ref{fig:massofstars} with and without feedback. In all cases, feedback reduces the mass of stars formed. We also see that the mass of stars forming slows down with time. In the lowest mass model with feedback, M1R1FB, star formation has largely ceased by around 7 Myr, though in the other models star formation is ongoing. We see that whilst star formation ceases in model M1R1FB, star formation continues throughout the duration of the equivalent model with no feedback, M1R1. This is because in the model with feedback, gas has been mostly dispersed by 7 Myr and is low density, whereas without feedback, there is still quite a lot of dense gas in and around the central clusters. Consequently after 20 Myr, the mass of stars formed in the model with feedback is around half that of the model without feedback. The reduction in mass with feedback included is lower in the other models, however these run for shorter times. If we compare at the same time (e.g. 5 Myr) then the difference with and without feedback is fairly similar for all the models. This suggests that  the main difference with the models is the timescales over which star formation occurs. For the higher mass models, star formation occurs over relatively short periods before ionisation has much effect. Whilst for the low mass models, there is opportunity for massive stars to form and significantly ionise the surrounding gas before further star formation occurs. Similarly to the results in Figures~\ref{fig:region1} and \ref{fig:region2} the ionisation is having a similar impact in the models at different times, but for the high mass models, large amounts of star formation has already occurred by later times. 

\begin{figure}
\centerline{\includegraphics[scale=0.42]{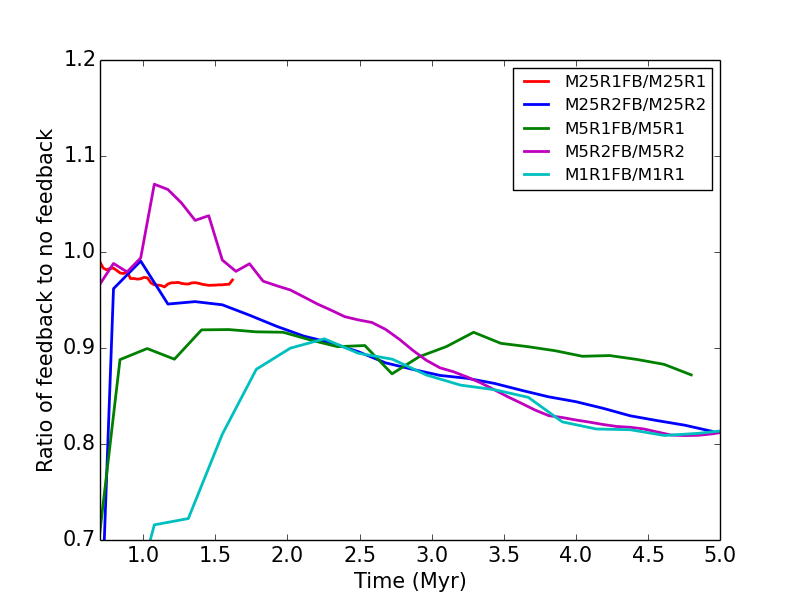}}
\caption{In this figure we plot the ratio of the mass of stars formed in each model with feedback, divided by the mass of stars formed without feedback at the equivalent time. For clarity, only part of the whole timescale from Figure~\ref{fig:massofstars} is shown.}   
\label{fig:massratio}
\end{figure}

We further compare the reduction in the mass of stars formed with and without feedback in Figure~\ref{fig:massratio}, where we plot the mass of stars in the model with feedback divided by the mass of the stars in the equivalent model without feedback, versus time. This figure supports the impression from Figure~\ref{fig:massofstars} that the addition of feedback does not significantly differ between the different models at a given time frame, although for the low mass models, feedback has an impact straight after star formation, possibly having a larger effect on dense gas around new sink particles compared to in the higher mass cases (in the M5R2FB model the ionization is actually having a slight net triggering effect at early times, though there has been relatively little star formation at this point).

\subsection{Stellar clusters}
\subsubsection{Clusters, OB associations, and the approach used here}
In the next sections we look at the evolution of clusters based on i) identifying clusters at different time frames (Section~\ref{sec:cluster_time}), and ii) identifying clusters at one time frame and following the constisuent sink particles over time (Section~\ref{sec:partevol}). For i) we refer to these simply as clusters. For ii) we identify objects where the sink particles stay clustered together and where they disperse, grouping these into clusters which remain or which disperse. For identifying clusters, we simply use a clustering algorithm, as typically used by observers (e.g. \citealt{2022A&A...661A.118C}, \citealt{2021A&A...646A.104H}, \citealt{2020A&A...635A..45C}, \citealt{2019ApJS..245...32L},  \citealt{2017A&A...600A.106C}). As such the clusters do not necessarily need to be gravitationally bound.

OB associations are groups of stars which are thought to originate from the same star formation event but are spread out on the sky. In their review,
\citet{Wright2022} suggest that densities of Galactic OB associations are typically  $0.001-0.1$ M$_{\odot}$ pc$^{-3}$, based on their typical sizes and masses.
Although the OB associations listed in \citet{PZ2010} tend to be fairly compact, other authors identify OB associations as spatially larger. The Galactic OB associations in \citet{Wright2020} can be 100 pc or more in size, and within them contain open clusters or smaller OB associations, whilst
OB associations listed in external galaxies are typically $\sim$ 80 pc (e.g. \citealt{Elmegreen1999}).
We simply compare the properties of our star forming regions with characteristics of Galactic OB associations using \citet{Wright2020} as a basis (Section~\ref{sec:OBassociations}).

\subsubsection{Evolution of clusters}
\label{sec:cluster_time}
\begin{figure*}
\includegraphics[bb=0 100 600 830, scale=0.65]{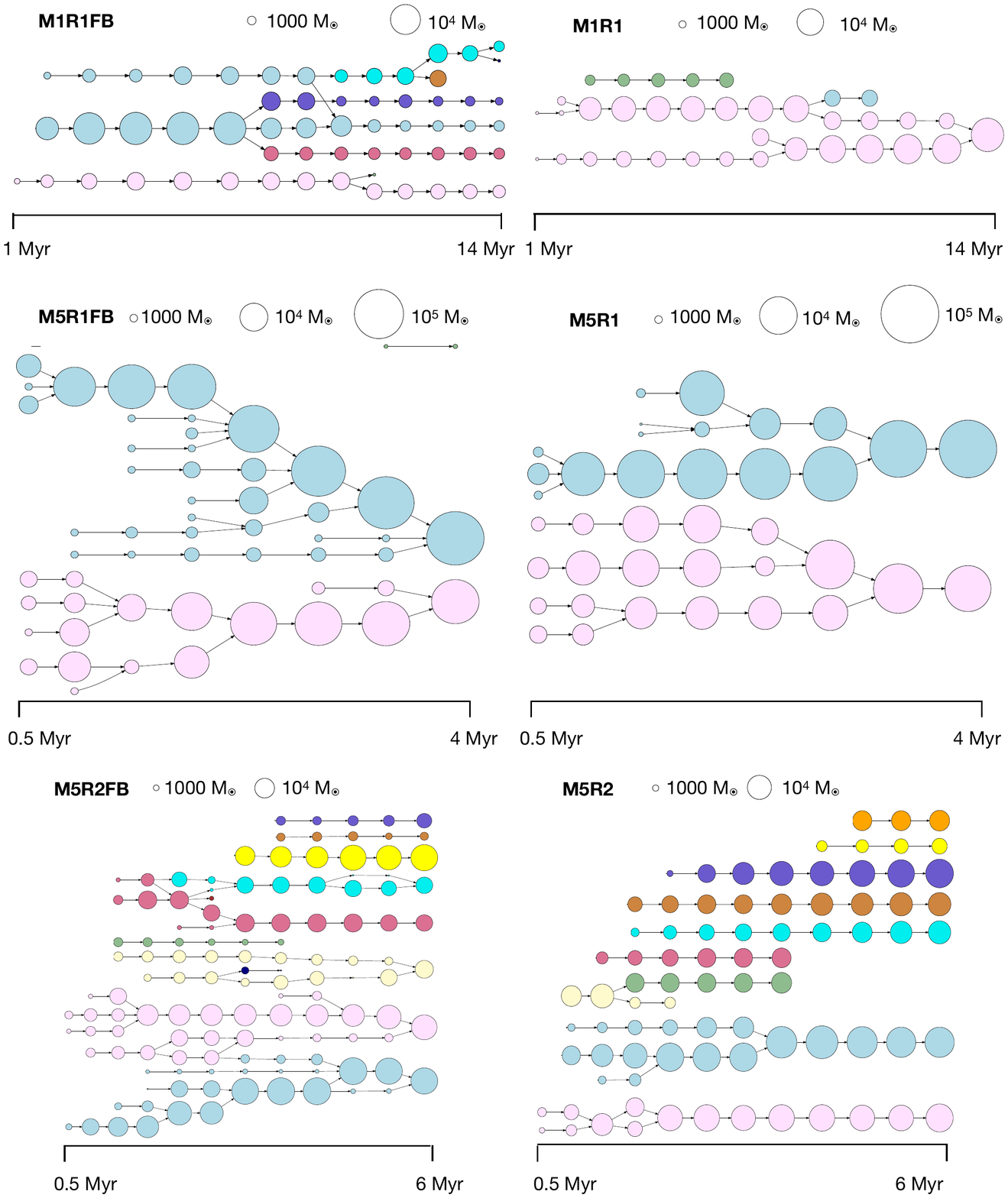}
\includegraphics[bb=0 540 600 820, scale=0.65]{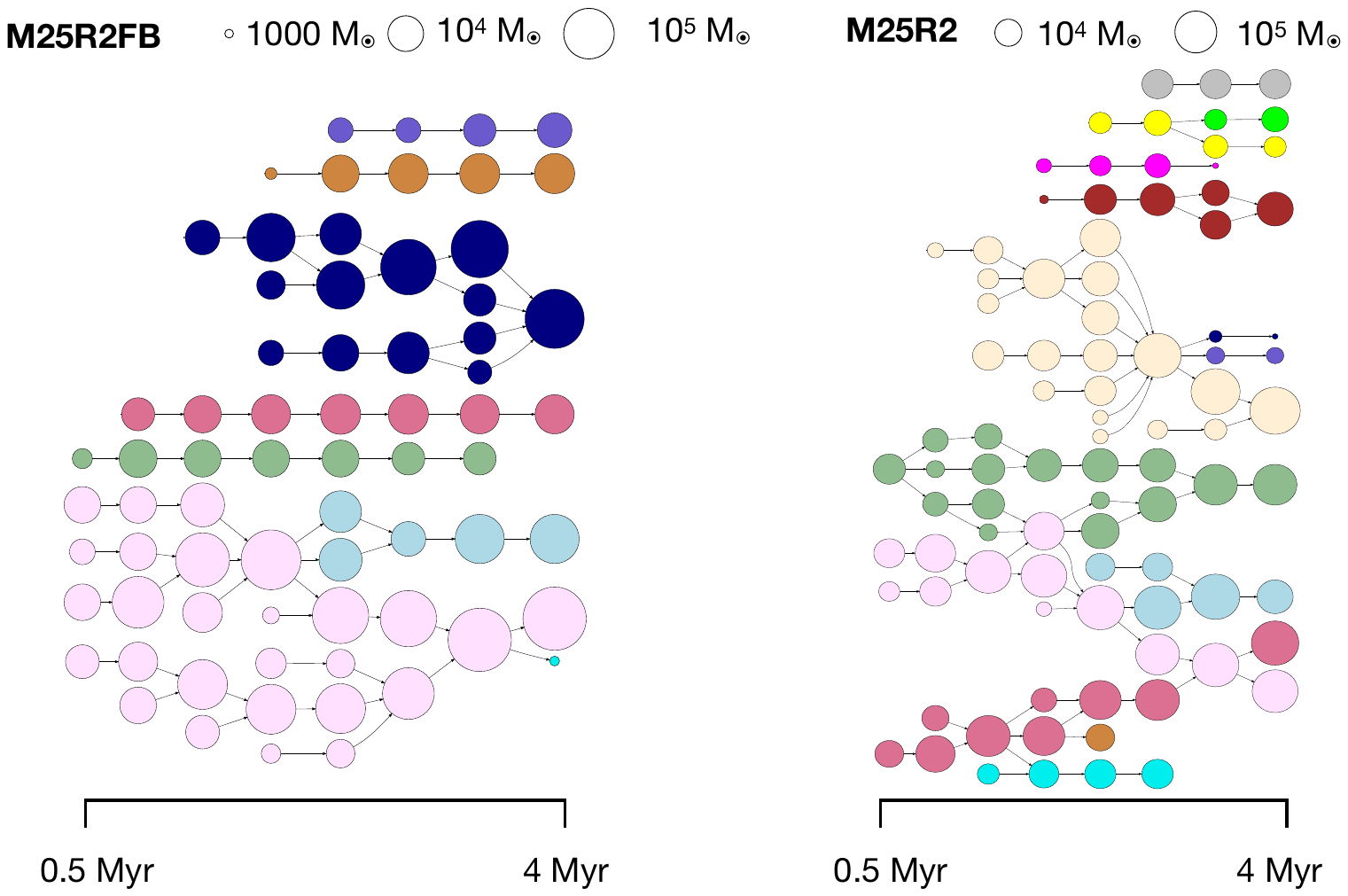}
\caption{Cluster merger trees are shown for the different models, with time going from left to right. The size of the points scales with the log of the mass. Each final cluster, whether at the last time shown or earlier, has an individual colour. In some cases there is a choice of colours when clusters split, this is just made randomly.}
\label{fig:mergertree}
\end{figure*}

Our FoF algorithm is similar to the density-based clustering algorithm DBSCAN \citep{Ester1996} but slightly simpler and easier to automate. Sinks are selected which are within a given distance of each other and are assigned into groups. We choose a distance of 2 pc, which is larger than the value of 0.5 pc used in \citet{Dobbs2022}, for a few reasons. Firstly this identified clusters which would be picked out by eye across the different data sets. Secondly, when following the clusters over time their relative size, shape, positions and memberships are liable to changes whilst the number of sinks unassociated with clusters increases (see e.g. Fig.\,\ref{fig:region1}). With this distance the algorithm is able to correctly identify a cluster at an earlier time and then find the same cluster (as would be picked out by eye) at later times, after the changes. Thirdly we found that 2 pc was reasonable using the approach of \citet{Rahmah2016} to find the optimal distance scale of a dataset.  Their approach works by calculating the distance from every star to its $N^{\rm th}$ nearest neighbour (here, $N=20$), sorting these distances from lowest to highest. Plotting order number vs distance, the scale is defined as the distance at which the maximum change in gradient of the slope occurs (see also \citealt{Buckner2022}). We checked the results of our algorithm with those obtained using DBSCAN, finding them to be very similar and often the clusters identified are identical. 

When deciding the best algorithum to use we also considered HDBSCAN \citep{HDBSCAN_ref}, a hierarchical implementation of DBSCAN better at finding clusters in varying density datasets. Unlike the original implementation, HDBSCAN does not require the user to specify a distance scale for a dataset, only the minimum cluster size (number of members). We found it produced clusters which agreed well with what would be picked out by eye, but as the optimum minimum cluster size needed to be found for each dataset it wasn't ideal for our intended purpose. Furthermore in some instances additional settings (e.g. a minimum distance between clusters for them to be considered separate) were needed to produce sensisble results, eliminating the advantage of the algorithm that it does not require user-defined distance scales. As such, we decided against using HDBSCAN in favour of our FoF algorithm.

To be considered a cluster by our FoF algorithm, we required spatial groupings to have more than 10 sink particles, although we reduced this to 7 for the M1R1 run because there were so few sink particles in this model. In all models, the number of sinks was larger (but their mass smaller) in the models with feedback. 
We show the evolution of the clusters as cluster merger trees (see also \citealt{Guszejnov2022}), similar to galaxy merger trees, in Figure~\ref{fig:mergertree}.
We follow the evolution of clusters over time, by running our algorithm at different timeframes, and identifying which particles lie in the clusters at different times (as seen in the appendix, the clusters' evolution also makes sense by eye). In Section~\ref{sec:partevol}, we take an alternative approach, where we take clusters at a particular time, and follow what happens to the constituent sink particles over time.  

Figure~\ref{fig:mergertree} shows that mergers and splitting of clusters (as denoted by the clustering algorithm picking up multiple versus single clusters at adjacent times) is commonplace in all models. There is also a tendency for more mergers and splits to occur in the models with feedback. As shown in \citet{Dobbs2022}, the massive clusters tend to form via the merger of a number of smaller clusters. The lower mass clusters, unless they have recently split from more massive clusters, tend to be clusters which have formed with a lower mass and stayed low mass for the duration of the simulation. There are more mergers in the simulations with higher gas masses, here gravity will be stronger and driving additional mergers, as well as promoting the formation of more clusters. As inidicated by the figures in Appendix~\ref{sec:appendix1}, clusters are not spherically symmetric (though they become more so with time), and sinks in the outer parts can be more loosely associated and may be unbound from the rest of the cluster and later split. Groups of stars which are not bound to the cluster could arise because they are spatially coincident, but not bound. Either they simply form close to the cluster, or dynamically come close together, but are not bound. 

The end cluster masses (at least for the more massive clusters) mostly tend to be similar with and without feedback, which is perhaps not surprising when the total stellar masses are more similar (Figure~\ref{fig:mergertree}). The exception is the lowest gas mass models, M1R1FB and M1R1. At the end point shown in Figure~\ref{fig:mergertree}, the model without feedback, M1R1 produces a cluster which is an order of magnitude more massive than any of the clusters which occur with feedback. We also see that typically there are more clusters in the model with feedback, and for example after 15 Myr, with no feedback there is only one cluster compared to typically three with feedback. Even allowing for the caveat that the sink particle numbers are different, by eye there appear to be more clusters in the model with feedback at later times, whereas in the model with no feedback, the sinks all merge into the cluster seen at $x=3.95$ and $y=4.315$ in Figure~\ref{fig:region1} at the 7 Myr time (see also Figures~\ref{fig:app1} and \ref{fig:app2} in the appendix). Even though at earlier times in model M1R1FB there is a more massive cluster, this is fairly asymmetric, and at least at 5 and 6 Myr it is bipartite, and simply splits apart by 7 Myr.

Some clusters cannot be followed to the final time, sometimes these are ones which occur as a result of a split and disperse. In addition, in some of the no feedback models, the number of particles in the clusters is low which artifically leads to cluster dispersal (see Section~\ref{sec:partevol}).

\begin{figure}
\includegraphics[scale=0.43]{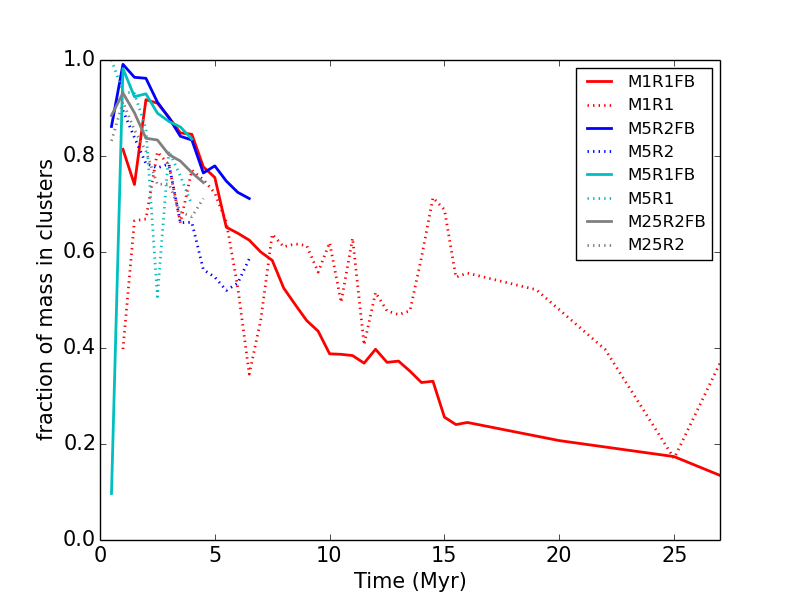}
\caption{The fraction of stars (by mass) which lie in clusters is shown versus time for the different models.}
\label{fig:fraction}
\end{figure}

Whether stars form in clusters or not has been the subject of considerable debate \citep{Bressert2010,Ward2018,Ward2020,Grudic2021}. In Figure~\ref{fig:fraction} we show the fraction of stars, by mass, which lie in clusters with time for each model. All show a decreasing fraction of stars with time. The high fractions at early times suggest that stars do tend to form in clusters, but due to the dynamics, many are ejected. Some are only loosely associated with the cluster and simply not picked up by the nearest neighbour algorithm at later times. The low mass models are the only models which run to longer timescales. In these cases, the model with feedback (M1R1FB) stays around 20\% up to 40\,Myr (the model with no feedback is higher at $\sim30\%$). This is not dissimilar to the fiducial simulation of \citet{Grudic2021}, where the fraction is 10\%, though there they see a wide range of fractions of stars in bound clusters for their models. 

\subsubsection{Cluster properties}
\label{sec:properties}
\begin{figure}
\centerline{\includegraphics[scale=0.63]{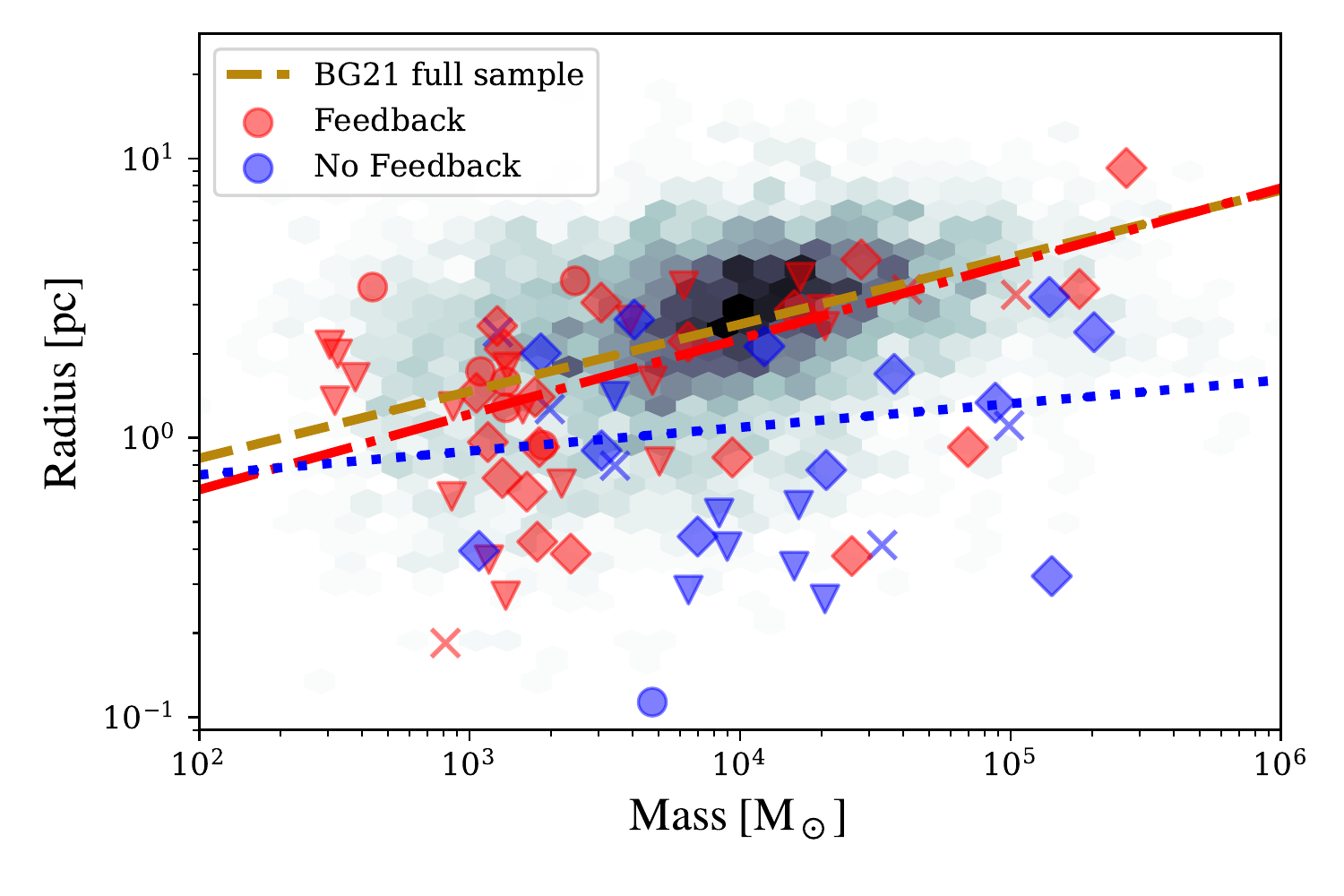}}
\caption{The radii and masses of the clusters found in the simulations are shown overplotted on observational data from \citet{Brown2021}. The circles represent clusters from the low mass M1R1 and M1R1FB models, the crosses from the medium mass Region 1 (M5R1 and M5R1FB), the triangles from the medium mass Region 2 simulations (M5R2 and M5R2FB) and the diamonds from the high mass Region 2 simulations (M25R2 and M25R2FB). The clusters from the simulations with feedback match the observations better than without feedback since the cluster radii are too small without feedback. The red and blue lines show best fitting lines to the simulated data (see text), the dashed gold line is the fitted related from \citet{Brown2021} to their entire sample.}
\label{fig:brownfigure}
\end{figure}

In Figure~~\ref{fig:brownfigure} we show the masses and radii of clusters formed in all the simulations except M25R1 and M25R1FB. We use the half mass radius, as observers use; in \citet{Dobbs2022}, we did not use the half mass radius as there were too few stars, but the larger spatial scale for our clustering algorithm here is probably a better choice and means that taking the half mass radius here is more practical\footnote{The results here with the half mass radius are similar to taking a lower spatial scale in \citet{Dobbs2022} with the full radius.}. The clusters are taken from the end points of the simulations except for the low mass simulations, where we take a time of 9 Myr, and are overplotted on observed clusters from \citet{Brown2021}. The observed clusters are taken from a sample of nearby galaxies (LEGUS). 

Figure~\ref{fig:brownfigure} shows that the clusters from the simulations with feedback match the observations much better than without feedback. We see that without feedback, the cluster radii are on average too small compared to the observed clusters. Figure~\ref{fig:brownfigure} confirms what we could see by eye in Figures~\ref{fig:region1} and \ref{fig:region2}, that the clusters are larger in size with feedback. This increase of radiues with feedback is expected, since as gas is dispersed away from the cluster, the potential changes, and the cluster will then expand \citep{Geyer2001,Pfalzner2011}. We see significant expulsion of gas from comparing the gas mass within the vicinity of the cluster with and without feedback (see Section~\ref{sec:gas}). We also examined whether the velocity dispersion of the stars contributed to expansion, but there was not a significant difference with and without feedback. We compared radii of clusters with and without feedback at different times, and found that they start to diverge when clusters were around 1-1.5 Myr old. 

We also see that there is a slight tendency for higher mass clusters to have larger radii. \citet{Brown2021} find that the radii of the clusters over their full sample vary with the mass according to 
\begin{equation}
R=2.55\bigg(\frac{M}{10^4}\bigg)^{0.24}.
\end{equation}
For our models with feedback, we find a relation of 
\begin{equation}
R=2.26\bigg(\frac{M}{10^4}\bigg)^{0.27}, 
\end{equation}
in excellent agreement with the data. By contrast, without feedback, we find a relation of 
\begin{equation}
R=1.09\bigg(\frac{M}{10^4}\bigg)^{0.085}.
\end{equation}

Compared to the observations, the simulated clusters still have a slightly larger spread than the observations in mass and radii, even with the better matching clusters from the simulations with feedback. The higher masses could reflect that probably the star formation rates are high as in reality there would likely be a lower efficiency for star formation when we form sink particles. The observations are also not complete, particularly at lower masses and radii.  Consquently we cannot compare the relative numbers of different mass clusters.

\begin{figure}
\centerline{\includegraphics[scale=0.45]{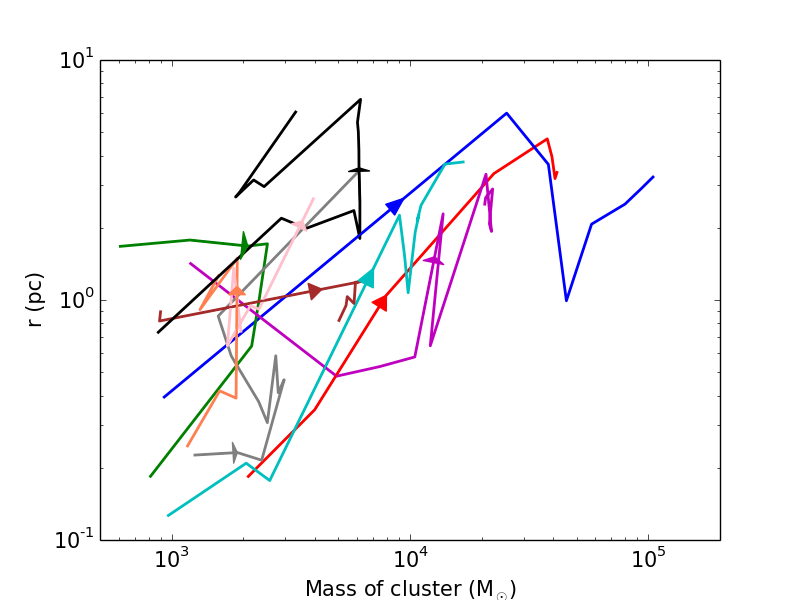}}
\caption{The mass and radius of ten clusters from the M1R1FB, M5R1FB and M5R2FB simulations (i.e. all with feedback) are shown where the lines represent time evolution (over periods of 3 to 8 Myr, in 0.5 Myr intervals). Most clusters exhibit evolution from lower left to top right, but some just increase in radius (including the black line, the only cluster from the low mass simulation M1R1FB shown), and the cluster shown in green decreases in mass and radius due to splitting. Spikes tend to indicate when clusters merge or split and take 2 or 3 time increments to settle.}
\label{fig:massradevol}
\end{figure}

In Figure~\ref{fig:massradevol} we take ten clusters from the different simulations which we are able to follow for a significant fraction of the simulation (and whose evolution is independent from each other, so for example not the result of splitting from another cluster which is shown), and show how their mass and radius evolves over time. All clusters are from the simulations with feedback. Taking clusters where we can follow the evolution biases us more towards clusters which end up relatively massive. As expected, most clusters show a tendency to increase in mass and radius and thus move diagonally upwards across the plot. The observed relation between mass and radius likely reflects this typical evolution. The increase in mass comes from additional star formation and mergers of clusters whilst the increase in radius likely comes from the clusters tending to be virialised, so that the radius increases as they grow in mass, plus the effect of feedback increasing the radius. 

However not all clusters follow a path of increasing mass and radius. We see some, e.g. grey, coral lines, where the radius increases but the cluster does not experience a significant increase in mass. We would expect some clusters to behave like this for a number of reasons, to fill the parameter space of observed clusters with smaller masses and larger radii, and because we know in the Galaxy OB associations exist which have large radii and small masses (the black line is from the M1R1FB model). 
The cluster represented by the green line shows quite different behaviour. This radius of this cluster decreases with time. This is the result of a cluster splitting into two subclusters - the cluster grows by attaining more stars or merging with another group of stars, but then splits apart.

We also looked at the kinetic and gravitational energy (see also \citealt{Dobbs2022}, but found that the ratio of kinetic to gravitational energy was strongly concentrated around 0.5 ($0.5 \pm 0.05$) indicating the clusters were virialised, and showed no particular trends. We found that the 1D expansion / contraction was a slightly better predictor of cluster evolution and we discuss this further in the next section.

\subsubsection{Evolution of constituent sink particles in a cluster}
\label{sec:partevol}
\begin{figure*}
\centerline{\includegraphics[scale=0.6]{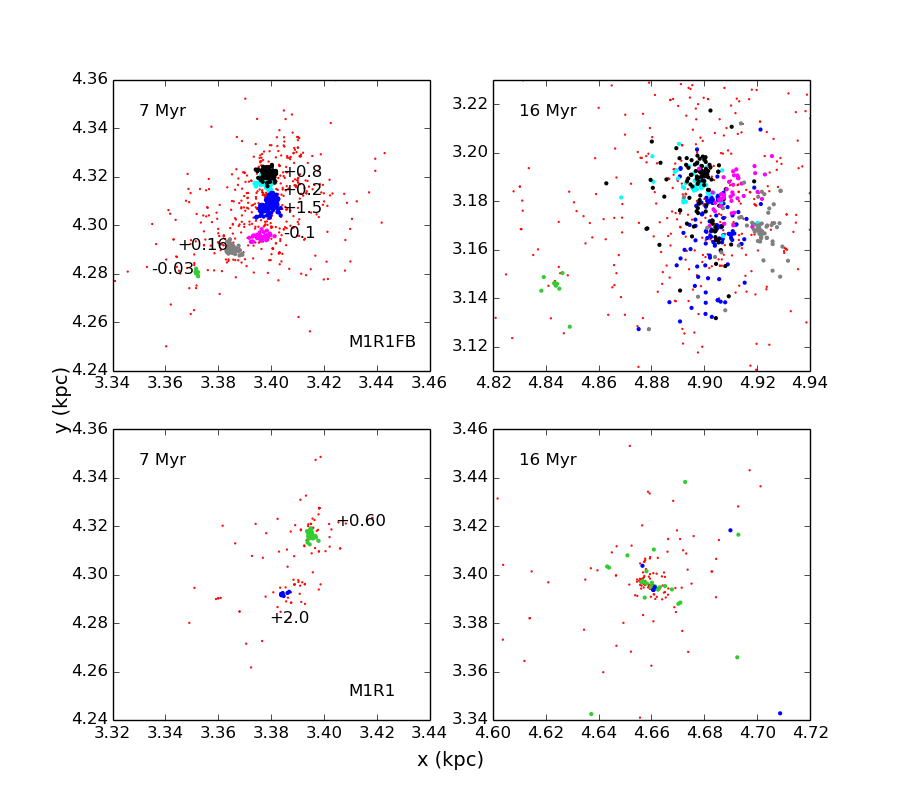}}
\caption{Clusters are shown from the low mass models at a time of 7 Myr (left), then the location of the constituent sink particles are shown on the right at 16 Myr. The red points indicate sinks which are not part of a cluster. For the feedback model (top) all the clusters show some expansion, though in many cases some core of sink particles remains. With no feedback, the two clusters simply merge though continuing star formation means some new stars are also present at 16 Myr. The numbers indicate the expansion / contraction velocity (see text), positive indicating expansion, negative contraction, and are in km s$^{-1}$.} 
\label{fig:partevol1}
\end{figure*}

In this section, rather than identifying clusters at different times and determining which correspond to the same object (constituting a largely similar but not necessarily identical set of constituent sink particles), we simply take the particles which constitute clusters at one time and show them at a later time. This has the advantage that we can follow any clusters which disperse as well as bound clusters.
In Figure~\ref{fig:partevol1}  we show clusters from the low mass region 1 simulations, M1R5 and M1R5FB, at a time of 7 Myr, and the locations of these particles at a time of 16 Myr. We also show the average expansion / contraction velocity calculated over all particles in the clusters at 7 Myr. This velocity is calculated as \citep{Kuhn2019,Buckner2022}
\begin{equation}
v_{\rm out}=\bf{v}.\bf{\hat{r}}
\end{equation}
where $\bf{v}$ and $\bf{\hat{r}}$ are calculated relative to the centre of mass of the cluster.

We show all 6 clusters found from the nearest neighbour algorithm at 7 Myr for the M1R5FB model. At 16 Myr, it is evident that in the model with feedback (M1R1FB) these clusters have expanded quite a lot. The clusters which are picked out by the algorithm at 16 Myr tend to be the cores of the distributions highlighted at 7 Myr, plus possibly stars previously identified in another cluster. Many stars of the `original' clusters are now at larger radii from these cores. 

The clusters shown in magenta and blue have more or less dispersed at 16 Myr. We find poorly resolved clusters ($<20$ particles) tend to disperse, these are stated explicitly in the text as poorly resolved. By the end of the simulation (40 Myr, not shown), there are three clusters picked out with the algorithm. Two contain constituent stars from the black, cyan and blue clusters at 7 Myr. A third contains constituent stars from the cluster shown in grey. The clusters shown in green (which is poorly resolved) and magenta (which is well resolved) at 7 Myr have dispersed, i.e. the clusters picked out at 40 Myr don't contain any of these sink particles. 

The velocities for the clusters in Figure~\ref{fig:partevol1} (top panel) are all fairly low, and most are positive or zero, indicating expansion or no evolution. The cluster shown in blue has the largest expansion velocity, and by eye this cluster does appear to cover the greatest spatial extent at 16 Myr. Nevertheless some sink particles from this cluster still belong to clusters identified at 16 Myr, and as discussed above even 40 Myr.

In the lower panels of Figure~\ref{fig:partevol1}, we show clusters at the same times from the model with no feedback (M1R1). Here only two clusters are picked out. The cluster in blue has a high positive velocity but is poorly resolved, and is dispersed after 16 Myr, though by this time it has actually collided with the cluster in green. The one cluster picked out at 16 Myr is comprised of some sinks from the 7 Myr clusters, but also newly formed sinks, since as we show in Figure~\ref{fig:massofstars}, star formation continues in the no feedback case, unlike the feedback model. The cluster in green (well resolved) remains more intact, and has a lower expansion velocity, and also lower virial parameter. 

\begin{table}
\begin{tabular}{c|c|c}
\hline 
Model & No. clusters which  & No. clusters \\
& remain in some form & which completely disperse \\
\hline
M1R1FB & 4 & 1 \\
M1R1 & 1 & 0 \\
M5R2FB & 7 & 0 \\
M5R2 & 5 & 0 \\
M5R1FB & 4 & 4 \\
M5R1 & 2 & 4 \\
\hline
\end{tabular}
\caption{Table showing the resultant evolution of clusters in the different simulations, as measured by the eventual positions of the particles which make up clusters at an earlier time. Clusters where just some core remains (more typical in the M1R1FB model) are listed as `remain in some form'. The table only includes clusters with $>20$ sink particles.}
\label{tab:cluster_evolution}
\end{table}

\begin{figure*}
\centerline{\includegraphics[scale=0.6]{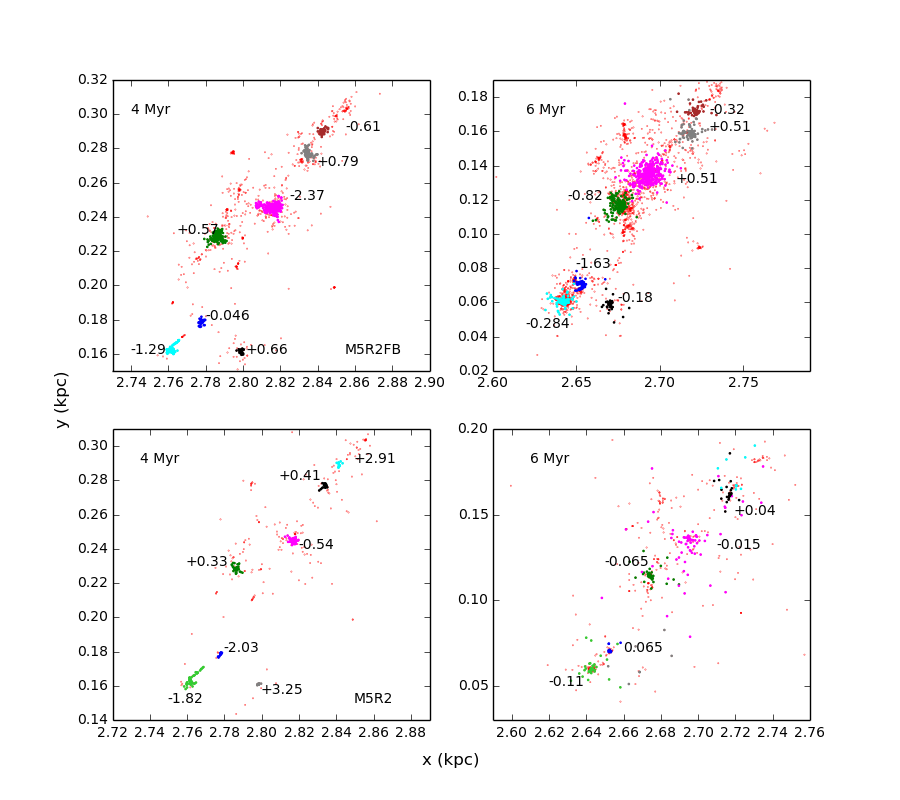}}
\caption{Clusters are shown from the medium mass region 2 models at a time of 4 Myr (left), then the location of the constituent sink particles are shown on the right at 6 Myr. The numbers indicate the expansion / contraction velocity (see text), positive indicating expansion, negative contraction, and are in km s$^{-1}$. Generally the clusters show less expansion compared to the low mass models (Figure~\ref{fig:partevol1}), and the velocities also include negative values indicative of contraction rather than expansion.}
\label{fig:partevol2}
\end{figure*}

In Figure~\ref{fig:partevol2}, we show clusters from the M5R2 and M5R2FB runs. Because there are many more clusters in M5R2FB, for clarity, we only study clusters with masses greater than $2\times 10^3$ M$_{\odot}$ which means that the selected clusters have more particles (at least 39). For the M5R2 model, all clusters are $>2\times 10^3$ M$_{\odot}$ and the minimum number of sink particles is 11. Although the clusters in the feedback model are spatially larger than without feedback, we otherwise see much greater similarity between the clusters selected with and without feedback compared to in the lower mass M5R1 models. At 4 Myr (left panels) the same clusters are selected in each simulation. Thus feedback appears to be having less effect on the clusters, as is also the case for the other higher mass simulations. 

We also see from the velocities, that there are more negative values indicative of contraction compared to the lower mass models. In the M5R1FB model, the velocities are all positive or only borderline negative. Most of the clusters are readily identifiable at the 6 Myr timeframe, and most are still fairly compact. We see in the feedback model that some of the clusters exhibit some expansion (e.g. green, grey), others appear very similar (e.g. cyan, brown) which mostly exhibit negative velocities at both time frames. The cluster shown in grey has positive velocities at both times, and shows more evident dispersal. In the model without feedback (lower panels), the velocities appear a fairly good indicator of short term cluster evolution.  The clusters shown in green and blue have high negative velocities and visually appear to contract slightly. The clusters shown in grey and cyan have high positive velocities, are poorly resolved and disperse rapidly, but the velocities are in agreement with the evolution.

We also performed the same analysis on clusters from the M5R1FB and M5R1 models at timeframes of 2 and 4 Myr though we do not show the plots. Between 2 and 4 Myr, nearly all the clusters picked out undergo collisions or interactions with each other. Hence the velocity analysis is not very meaningful because the clusters do not evolve in isolation at all (though to some extent this is also true of M1R1FB). Unlike the Region 2 models, we do see some examples of clusters identified at 2 Myr where by 4 Myr the constituent sink particles are fairly widely dispersed (and certainly no longer identifiable as clusters). However due to the compression of the gas, the clusters all end up quite close together, and some constituent sinks may have joined another cluster. There is some suggestion of similar behaviour to M1R1FB, whereby the region is compressed together, the clusters form and merge or collide, and in the resulting stellar population, some clusters remain as clusters, whereas some are or have dispersed. For model M5R2FB, which does not have such strong dynamics, and the clusters interact less, the clusters seem to remain intact at least for the duration we study them for. We also show the evolution of clusters from the different models in Table~\ref{tab:cluster_evolution}, indicating only those that are well resolved. Only the more dynamic, Region 1 simulations contain clusters which disperse.

\subsubsection{OB associations}
\label{sec:OBassociations}
Given the small number of  OB associations in our Galaxy with detailed information, we simply compare with those listed in \citet{Wright2020} where we have information on mass, size, ages and number of subgroups.
For our higher mass simulations, the stars tend to form in fairly bound clusters, roughly at similar times and are atypical of nearby OB associatons. The M1R1FB region however contains a lower stellar density ($\lesssim 0.1$ M$_{\odot}$ pc$^{-2}$), stars form over a prolonged time, and groups can be unbound themselves, and with respect to each other.

We compare the region in M1R1FB with specific regions given in \citet{Wright2020} in Table~\ref{tab:OBassociations}. As indicated in Table~\ref{tab:OBassociations}, some nearby OB regions are larger, and correspondingly tend to be older than M1R1FB, whilst some lesser known ones tend to be smaller and have fewer subgroups. Of the observed associations we have extensive information for, our simulated region is probably closest to Orion Ia. We show a visual comparison with Orion Ia in Appendix B. M1R1FB is similar mass and size to Orion Ia. The age spread is a little lower, Orion Ia is slightly smaller and also there is likely more gas still in the vicinity of Orion Ia. These could be related to the amount of feedback in the simulation, e.g. with less feedback, star formation may be more prolonged and larger volumes of gas remain. Prolonged gas inflow into the region could also lead to more gas present at later times.  Again with no feedback at all however (model M1R1) there is simply one cluster, which is not at all like Orion Ia, or indeed any of the other listed associations.   
\begin{table*}
\begin{threeparttable}
\begin{tabular}{c|c|c|c|c|c}
\hline 
Region & Mass  & Size & No.  & Age &Comparison  \\
 & (M$_{\odot}$) & (pc) & subregions & spread (Myr) & to M1R1FB \\
\hline
Sco OB2 & 4000 & $>100$ & 3 & $\sim 20$ & too large, star formation occurring over longer time \\
Orion Ia & 8500$\pm$1500 & $100$ & $\sim5$ & $\sim12$ & slightly bigger, slightly larger age spread \\
Vela OB2 & $>2300$ & - & 8 & 40 & older, more subgroups \\
Cygnus OB2 & 16500 & 200 & - & $\sim 7$ & too big \\
Perseus OB2 & 6000 & 40 & - & $>5$ & too small \\
Carina OB1 & $>2\times 10^4$ & $~100$\tnote{\textdagger} & 9 & - & too massive \\
Lacerta OB1 & 1 O star & - &  2 & few Myr & too small \\
\hline
M1R1FB & $10^4\pm1500$ & 70 & $\sim4$ & $\sim 7$ & - \\
\hline
\end{tabular}
\begin{tablenotes}
\item[\textdagger] \citet{Melnik2020}
\end{tablenotes}
\end{threeparttable}
\caption{Table showing observed OB associations \citep{Wright2020}, their properties and a comparison to the stellar distribution in model M1R1FB. The observed OB association which best resembles M1R1FB is Orion Ia (see Appendix~\ref{sec:appendix2}).}
\label{tab:OBassociations}
\end{table*}

\subsubsection{Evolution of gas}
\label{sec:gas}
\begin{figure*}
\centerline{\includegraphics[scale=0.74]{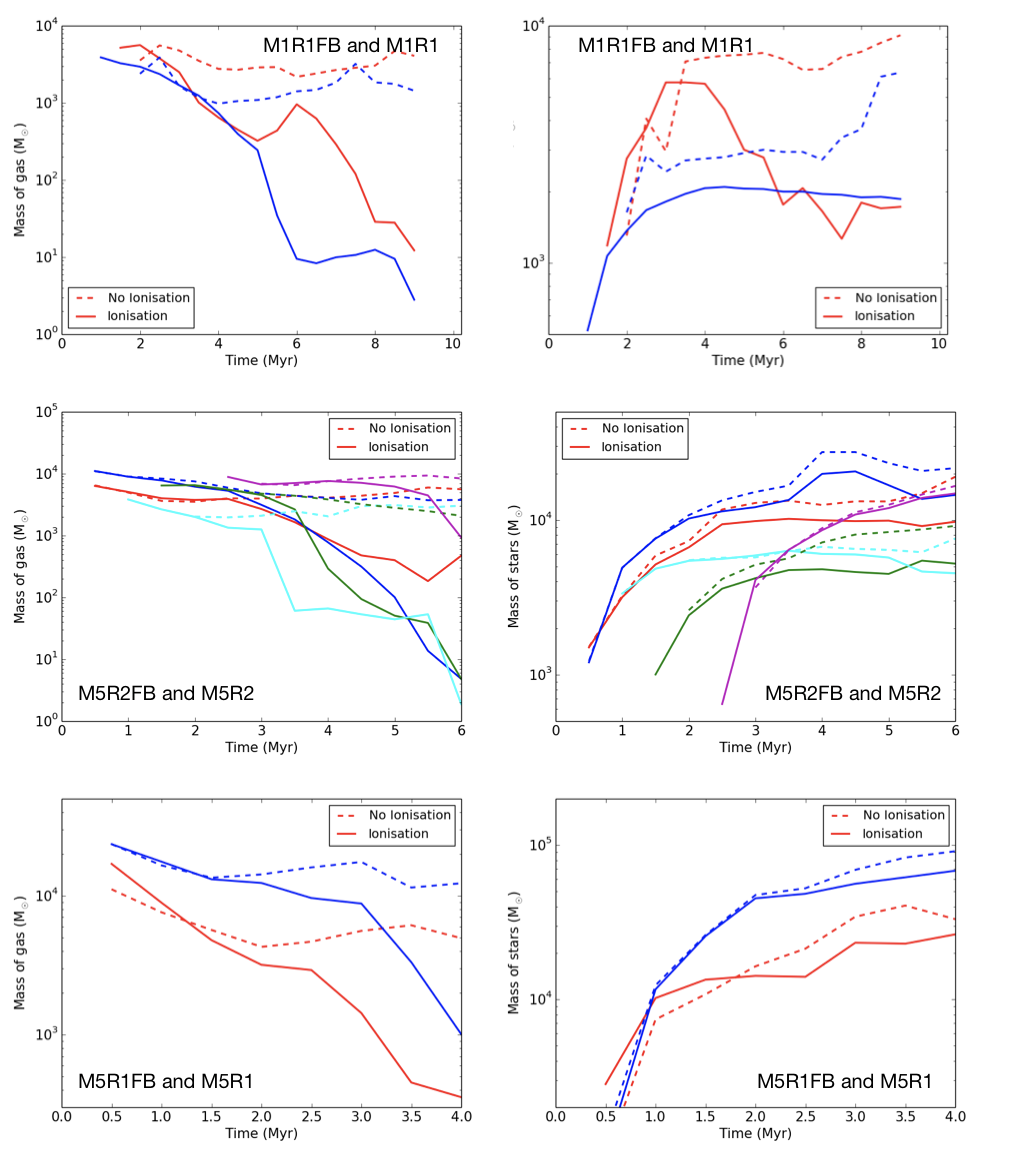}}
\caption{The mass of gas, and mass of stars within 5 pc of the centre of clusters is shown for the different models. Clusters are selected which have counterparts in the no feedback simulation, and the same colour lines are used to represent equivalent clusters.}
\label{fig:gasplot}
\end{figure*}
In this section we examine the evolution of gas, and also the stars, within a particular radius of the clusters in the simulations. This is to see how quickly / how much photoionization clears away the gas from clusters, and the effect on stellar mass. In \citet{Dobbs2022}, we chose radii of 1 and 2 pc. Here we choose a larger radius of 5 pc, partly because the ionisation has a stronger effect here on the surrounding gas, and also because the lower densities, lower amounts of star formation, and smaller number of clusters, means that in most cases (though not always) a radius of 5 pc will just contain one cluster. 

We show results for a sample of clusters from the M1R1 and M1R1FB (top), M5R1 and M51R1FB (middle) and M5R2 and M5R2FB simulations (lower panels) in Figure~\ref{fig:gasplot}, which correspondingly have the lowest to highest mass clusters forming.  We only show clusters which can be followed for most of the simulation, exclude other clusters which merge or split with those already shown, and only include those with a clear conterpart in the no feedback model. The mass of gas within 5 pc is shown in the left, and the mass of stars on the right. Lines of the same colour represent the same cluster identified from models with and without feedback. Recalling \citet{Dobbs2022}, where very massive clusters formed, there we saw a relatively small reduction in the stellar mass, up to around 25\%, whilst the gas mass could decrease by up to an order of magnitude. 

Starting with the low mass simulations (top panels) we see that the inclusion of ionisation has a dramatic effect on the mass of gas within 5 pc. We find that within 5 Myr (from 2 to 7 Myr) the mass of gas within 5 pc drops by 1 or 2 orders of magnitude for the four clusters when ionisation is included. By contrast, with no ionisation, the gas mass within 5 pc stays relatively constant.  The mass of stars within 5 pc is also substantially reduced, by a third or so at earlier times and more later, and again substantially more compared to \citet{Dobbs2022}.

In the M5R2FB  and M5R2 simulations (middle panel), the ionisation again has a clear effect on the gas, decreasing the gas mass by an order of magnitude and a half for the clusters shown in red, green and blue over 3 or 4 Myr. The effect on the cluster in magenta is much less, though this cluster evolves for less time. Again, for all the clusters with ionisation, the gas mass decreases over time, whereas it stays more or less constant with no ionisation. The stellar mass shows a more moderate change, unsurprisingly there is minimal change for the cluster shown in magenta, though for the cluster shown in red, the mass is less than half with ionisation. 

Finally the lower panels show the M5R1FB and M5R1 models, where higher mass clusters form, so closer to although still somewhat less massive than those in \citet{Dobbs2022}. Again ionisation has a clear effect on the gas mass. The gas mass decreases by a factor of around 2 to 4 over a timescale of 2 to 2.5 Myr. So this is less than the other simulations, but the timescale is shorter. The gas mass in the models without ionisation shows some small decrease, possibly because comparably more gas mass has been converted into stars. The ionisation reduces the mass by a factor of 2 for the clusters shown in red and green, but has minimal effect on the cluster shown in blue, which is also the most massive in any of the simulations. 

Overall there is a tendency for ionisation to have a greater effect on lower mass clusters. The trend is not completely clear because the timescales are longer for the lower masses. However more gas is removed and the stellar masses tend to be more reduced by ionisation after 3 Myr for the lower mass clusters compared to the higher mass clusters. Generally, gas is expelled from the vicinity of clusters on Myr timescales in agreement with observations. And, as discussed in Section~\ref{sec:sfr}, for the high mass clusters, stellar masses of $10^4$ or $10^5$ M$_{\odot}$ are accumulated before feedback has much effect, whereas for the lower mass clusters, mass is accumulated over a longer time and feedback has more chance to impact the cluster properties.

\subsection{Impact of supernovae}
\label{sec:supernovae}
\begin{figure*}
\centerline{\includegraphics[scale=0.65]{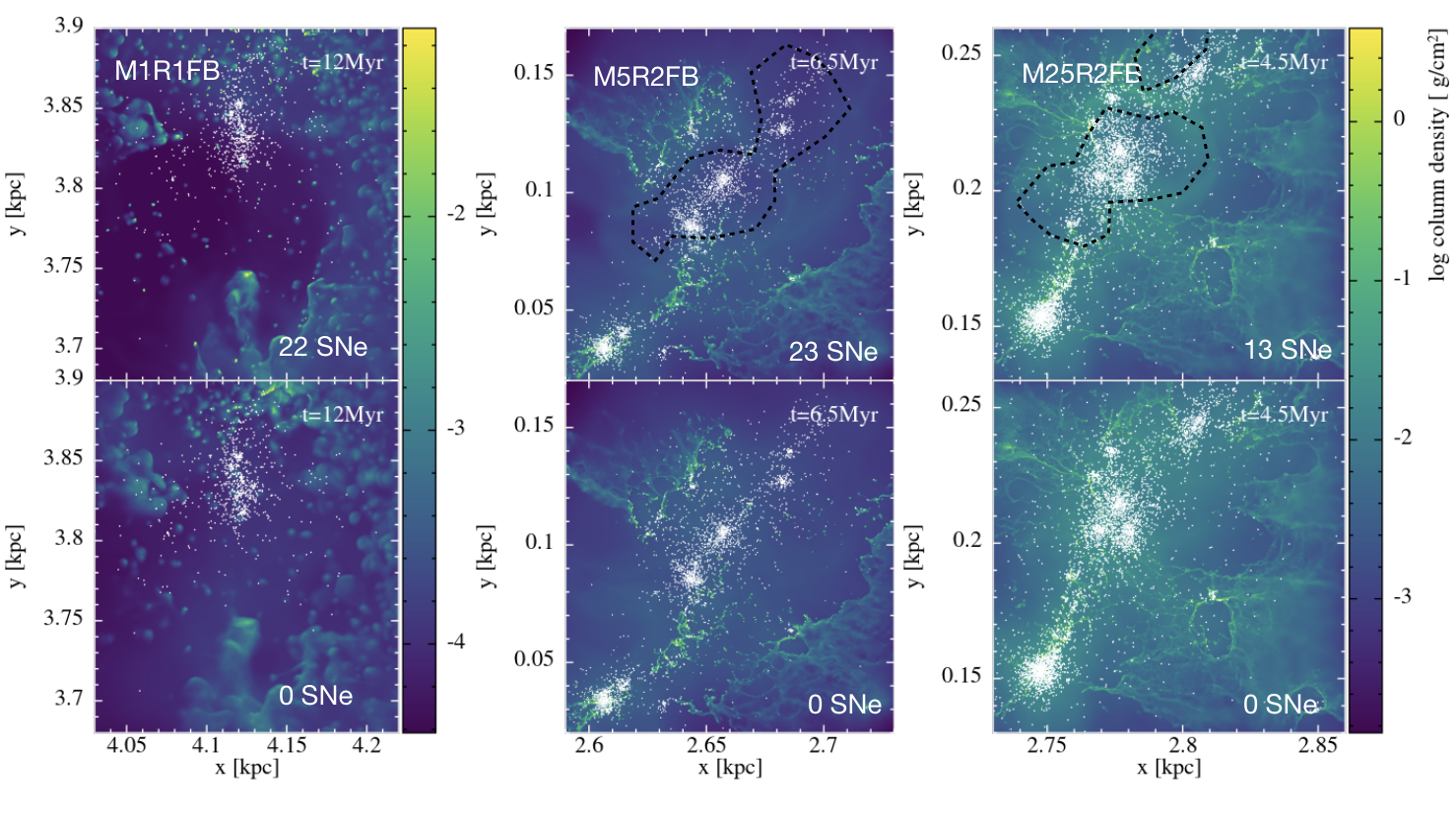}}
\caption{The upper panels show models with both ionization and supernovae, the lower panels there are no supernovae. The supernovae do not have a large effect, rather they fill already diffuse regions with hot gas. The dashed lines, indicate where the hot gas lies, in the left panels it fills the area shown with the lowest densities as seen by eye. The column density scale differs on the left hand panels, and the spatial extent is also larger.}
\label{fig:supernovae}
\end{figure*}
In the results presented so far, we include supernovae. We also repeated runs M1R1FB, M5R2FB and M25R2FB with only ionisation, starting from just before the first supernovae. Like \citet{Bending2022}, we did not switch ionisation off when stars were the age to undergo supernovae (see Herrington et al., in prep for this). We compare the M1R1FB, M5R2FB and M25R2FB models with and without supernovae in Figure~\ref{fig:supernovae}. The models are shown at the end points afor M5R2FB and M25R2FB, for M1R1FB, M1R1FB we show at 12 Myr, but the evolution of M1R1FB does not change significantly after 7 or 8 Myr, the gas simply becomes more diffuse. 

As seen in \citet{Bending2022}, the supernovae don't have a big impact and are secondary compared to the ionisation. The supernovae simply appear to fill the low density regions created by the ionisation with even lower density, hotter ($10^8$\,K) gas, marked explicitly for the M5R2FB and M25R2FB models. The supernovae appear to have a more marked effect in the lowest gas mass model, M1R1FB, which has had a similar number of supernovae to the M5R2FB model. The supernovae create a very clear cavity, and at the edges of this the gas is denser and more compressed compared to the case with only ionisation. The size of this cavity is also quite large, of order 100 pc.

The supernoave have a very minor impact on the star formation rate and clusters. The supernovae slightly increase the star formation rate compared to without supernovae; the density enhancement we see at the edges of the supernovae bubbles in Figure~\ref{fig:supernovae} can be enough to lead to slightly more stars. The supernovae can also slightly change the positions of the sinks, which changes the groupings of the sinks into clusters, but their spatial distribution is still fairly similar, and the overall properties and trends, e.g. the cluster radius relation do not change.

\section{Conclusions} 
We have performed simulations  of two sections along spiral arms with different initial gas densities, including photoionizing and supernovae feedback. The initial conditions are taken from previous galaxy scale simulations, and exhibit converging flows in spiral arms. One region exhibits strongly converging flows, and the other moderately converging flows. We change the initial gas mass to run lower gas density simulations whereby lower mass clusters form, to produce a population of clusters across a range of masses.

We find that photoionising feedback has a notable effect on cluster radii, and is required to produce the observed cluster radius mass relation (in agreement with \citealt{Hajime2022}). As photoionization clears away gas from the vicinity of the cluster, the gravitational potential is reduced and the cluster expands. Similarly, N-body simulations of clusters where gas is explicitly removed also show that the radius of the cluster thereafter increases \citep{Goodwin2006,Moeckel2010,Lughausen2012}.

Supernovae have little impact on cluster properties. Typically they occur after the cluster masses and radii have become established. In terms of cluster masses, spatial distribution and simply the number of clusters, we find that the effect of photoionization is much greater on these properties when lower mass clusters form, i.e. at lower gas densities and lower converging flows. This is because star formation is prolonged, so photoionization has time to act, whereas the high mass clusters form quickly with high star formation rates. The star formation rate, and mass of clusters formed, is suppressed more in the lower density regions (although the total stellar mass is only reduced by a factor $\lesssim 2$). This is similar to previous results studying individual molecular clouds \citep{Dale2012,Dale2017,Colin2013,Gavagnin2017,Ali2019,Kim2021b}. One caveat to this result is that we don't include radiation pressure, which is expected to be more relevant in high density regimes \citep{Krumholz2009,Kim2018}. IR radiation from dust is also expected to be relevant in high density dust rich regimes \citep{Skinner2015,Raskutti2016,Tsang2018} although very recent work by \citet{Menon2022} finds that radiation from dust only has a small effect on the star formation rate, and may not be that effective at dispersing the gas \citep{Tsang2018,Ali2021}.


During cluster evolution, mergers and splits are common, particularly in higher density, more dynamic regions. The most massive clusters which form in each simulation tend to be those which have undergone mergers. Some clusters move diagonally towards higher masses, and higher radii, in the cluster radius mass plot, as might be expected if clusters grow. However clusters can stay at, or even move to lower masses, simply because the cluster is relatively isolated, or where a cluster has split into lower mass clusters. The fraction of stars in clusters also decreases with time, as the clusters interact and stars are ejected, and through dynamical ejection. For our longest run model, the lowest mass model, this fraction flattens out around 20\%.

We only see a low density group of stars similar to an OB association form in our low mass model with feedback (M1R1FB). This is not surprising as this is the least gravitationally dominated, most dynamic model. Unlike \citet{Grudic2021}, who model a $10^7$ M$_{\odot}$ cloud, and obtain a dozen or so clusters in one association, our region is lower density, lower mass and produces a smaller number of groups, some of which themselves are more like associations. The star forming region here is not dissimilar to Orion Ia. The evolution of this region supports the findings from recent work \citep{Wright2016,Ward2018} that OB associations are not simply clusters which are expanding, but their evolution is more complex and there is no simple picture of uniform expansion.

\section*{Data Availability}

The data underlying this paper will be shared on reasonable request to the corresponding author.

\section*{Acknowledgments}
We thank the referee for a useful report which helped clarify some aspects of the paper. This work was performed using the DiRAC Data Intensive service at Leicester, operated by the University of Leicester IT Services, which forms part of the STFC DiRAC HPC Facility (www.dirac.ac.uk). The equipment was funded by BEIS capital funding via STFC capital grants ST/K000373/1 and ST/R002363/1 and STFC DiRAC Operations grant ST/R001014/1. DiRAC is part of the National e-Infrastructure.
Some of the figures in this paper were made using splash \citep{splash2007}. We also used the graphviz package to make Figure~\ref{fig:mergertree}.
CLD, TJRB and ASMB acknowledge funding from the European Research Council for the Horizon 2020 ERC consolidator grant project ICYBOB, grant number 818940. 
ARP acknowledges the support of The Japanese Society for the Promotion of Science (JSPS) KAKENHI grant for Early Career Scientists (20K14456). 

\bibliographystyle{mn2e}
\bibliography{Dobbs}
\bsp

\appendix
\section {Evolution of clusters for each model}
\label{sec:appendix1}
\begin{figure*}
\centerline{\includegraphics[bb=0 0 420 630, scale=1.04]{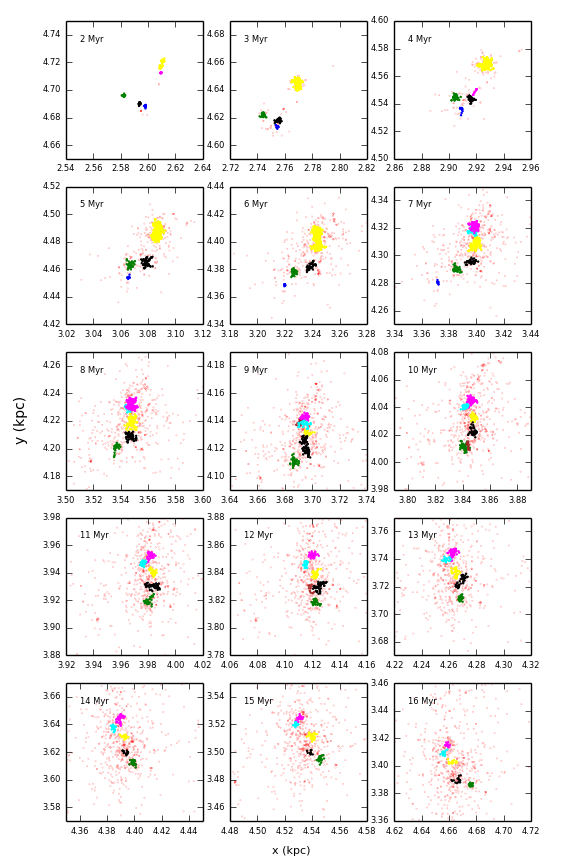}}
\caption{The clusters identified from the M1R1FB model are shown between 2 and 16 Myr. Points indicated in red are sink particles not identified as part of a cluster, these are shown with smaller points for clarity. At the end point of the simulation, 40 Myr (not shown), there are three clusters.}
\label{fig:app1}
\end{figure*}

\begin{figure*}
\centerline{\includegraphics[bb=0 0 420 630, scale=1.04]{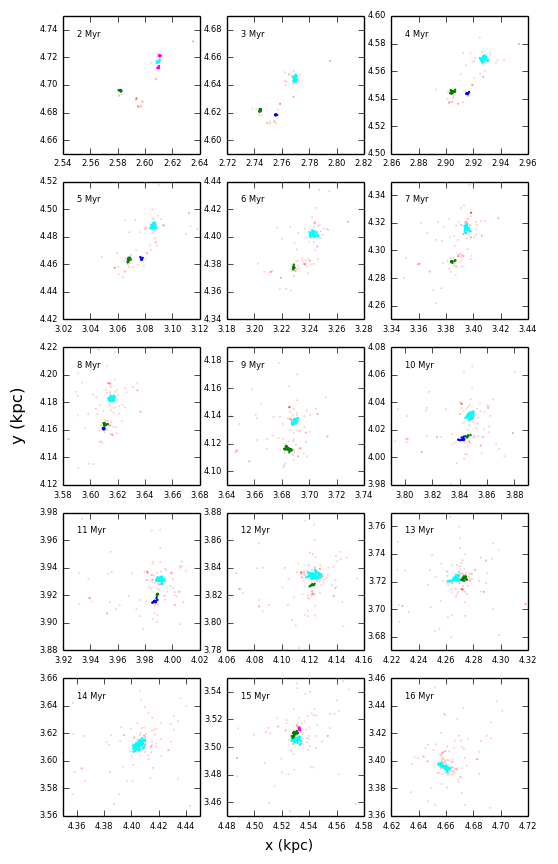}}
\caption{The clusters identified from the M1R1 model are shown between 2 and 16 Myr.}
\label{fig:app2}
\end{figure*}

\begin{figure*}
\centerline{\includegraphics[scale=0.94]{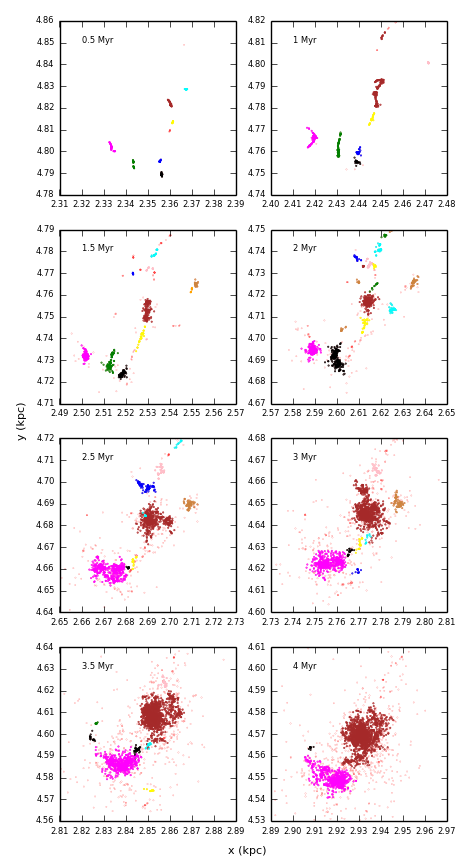}}
\caption{The clusters identified from the M5R1FB model are shown between 0.5 and 4 Myr.}
\label{fig:app3}
\end{figure*}

\begin{figure*}
\centerline{\includegraphics[scale=0.94]{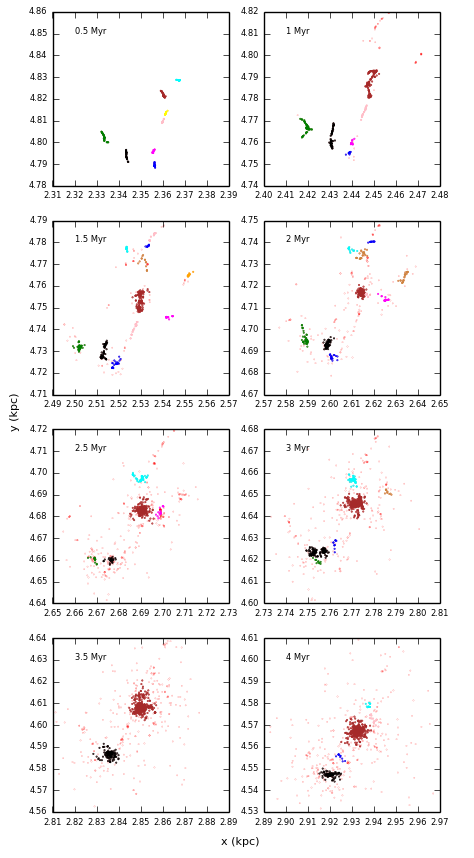}}
\caption{The clusters identified from the M5R1 model are shown between 0.5 and 4 Myr.}
\label{fig:app4}
\end{figure*}

\begin{figure*}
\centerline{\includegraphics[bb=0 20 800 480, scale=0.64]{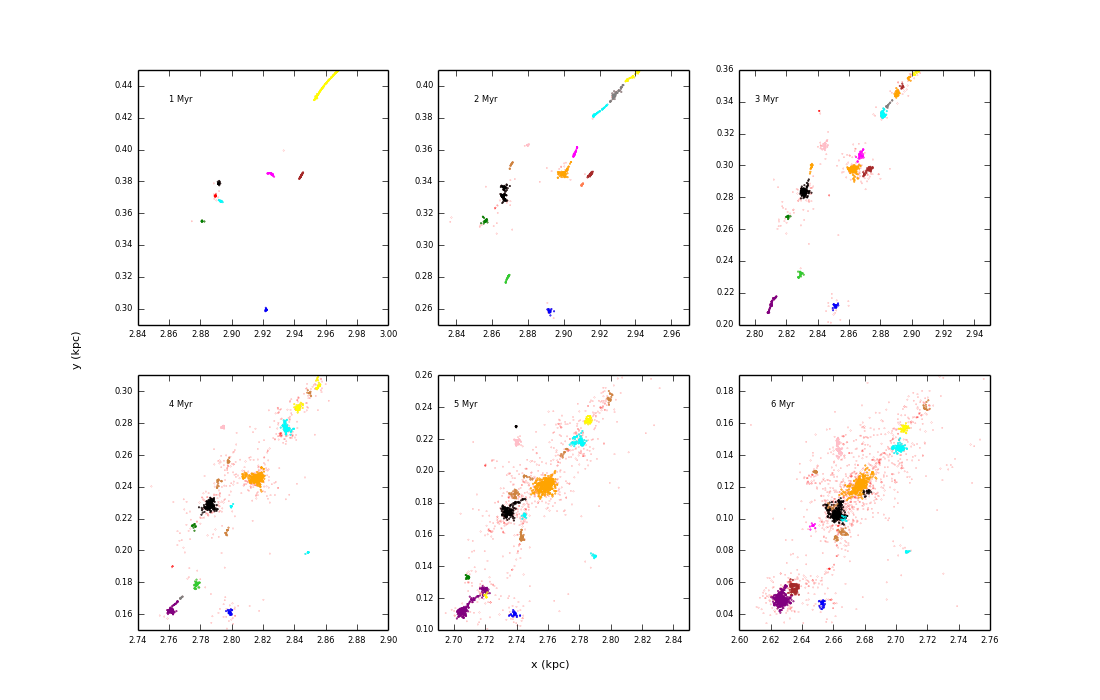}}
\caption{The clusters identified from the M5R2FB model are shown between 1 and 6 Myr.}
\label{fig:app5}
\end{figure*}

\begin{figure*}
\centerline{\includegraphics[bb=0 20 800 480, scale=0.64]{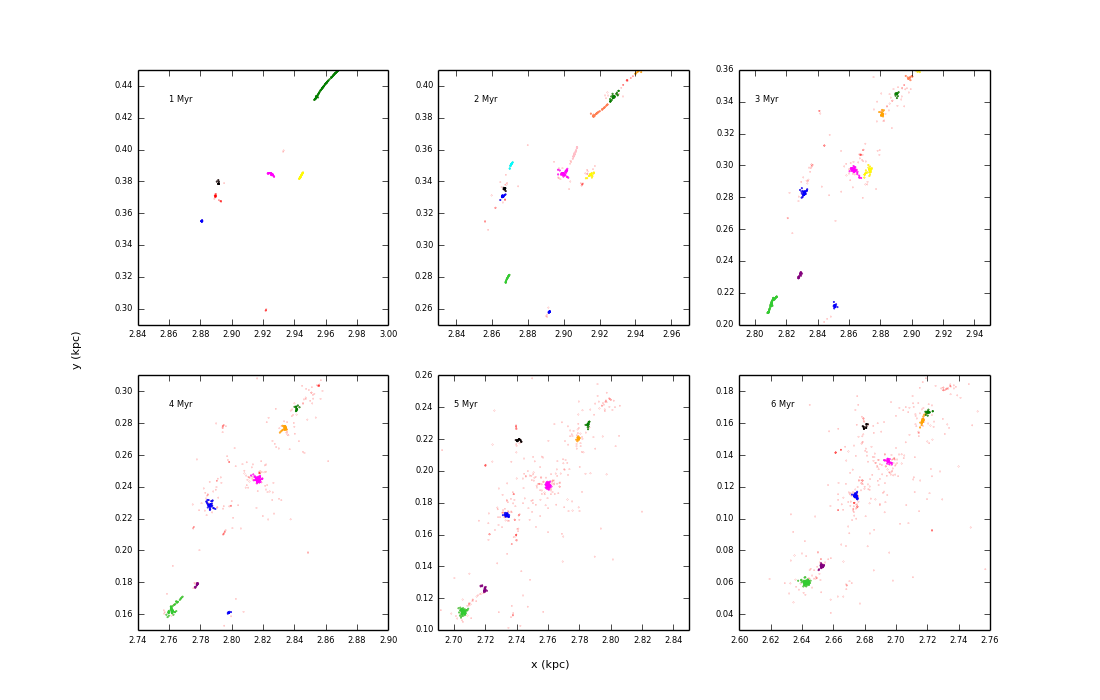}}
\caption{The clusters identified from the M5R2 model are shown between 1 and 6 Myr.}
\label{fig:app6}
\end{figure*}

In this appendix we show the clusters identified by our friends of friends algorithm over a series of timesteps for the M1R1, M1R1FB, M5R1, M5R1FB, M5R2 and M5R2FB models, i.e. all the medium and low mass models presented, and for which most of the analysis is focused on. In Figures~\ref{fig:app1} and \ref{fig:app2} we show the cluster evolution for M1R1FB and M1R1 respectively up to a time of 15.5, though we do follow the clusters for a time up to 40 Myr. In Figures~\ref{fig:app3} and \ref{fig:app4} we show the cluster evolution for M5R1FB and M5R1, and in Figures~\ref{fig:app5} and \ref{fig:app6} we show the cluster evolution for M5R2FB and M5R2.

\section {Comparison to Orion}
\label{sec:appendix2}
\begin{figure*}
\centerline{\includegraphics[scale=0.55]{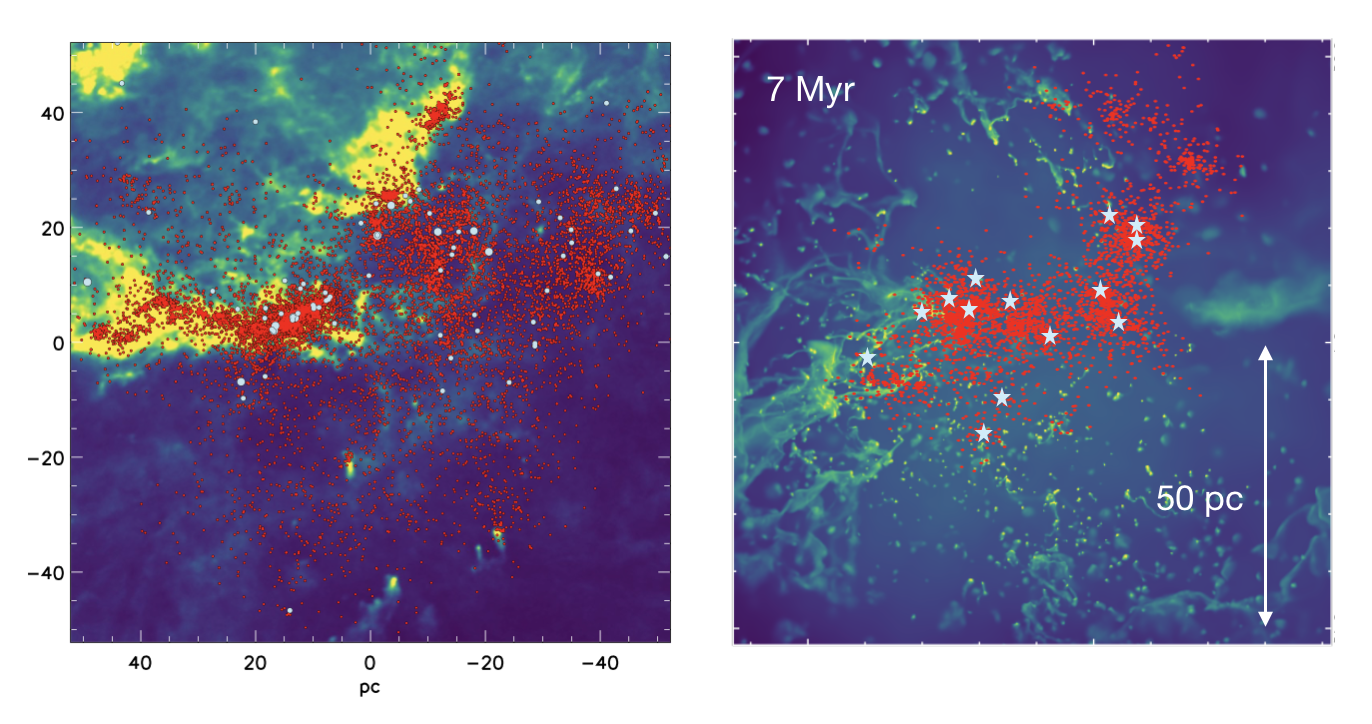}}
\caption{The left hand panel shows the Orion region \citep{Megeath2022}. The background is an extinction map from the Planck Legacy Archive. OB stars are indicated by light blue circles. The red points indicate stars, either from the Spitzer survey or GAIA DR2.  The right hand panel shows the simulation, the background green - blue colour scale is the gas column density. In this figure, sinks have been converted to stars according to the method in \citet{Liow2022}, with stars distributed within 10 pc of the original positions of the sinks. The main difference compared to just using the sink particles is that the number of points better matches the observations. O stars are indicated by light blue stars.}
\label{fig:Orion}
\end{figure*}

We show a visual comparison of our M1R1FB model with Orion in Figure~\ref{fig:Orion}. The simulation figure (right panel) shows the region in the $y-x$ plane (i.e. as would be face on). Although not shown, the region displays a similar morphology in terms of more gas at the lower part of the image, and the stars occupying a broad diagonal distribution to the top right, in the $z-y$ plane (i.e. as would be viewed through the plane of the disc in our Galaxy).

\label{lastpage}
\end{document}